\documentclass[12pt]{iopart}
\usepackage{iopams}
\usepackage{graphicx,float}
\usepackage{color}
\usepackage{subfigure}
\newcommand{\RR}{{R}}
\newcommand{\bc}{\begin{center}}
\newcommand{\ec}{\end{center}}
\newcommand{\be}{\begin{equation}}
\newcommand{\ee}{\end{equation}}
\newcommand{\bea}{\begin{eqnarray}}
\newcommand{\eea}{\end{eqnarray}}
\eqnobysec
\def\1.2{\frac{1}{2}}

\begin{document}
\title{Entanglement entropy in quantum impurity systems and systems with boundaries}
\author{Ian Affleck$^1$,  Nicolas Laflorencie$^2$ and Erik S. S{\o}rensen$^3$}
\address{$^1$Department of Physics \& Astronomy, University of
    British Columbia, Vancouver, B.C., Canada, V6T 1Z1}
\address{$^2$Laboratoire de Physique des Solides,
Universit\'e Paris-Sud, UMR-8502 CNRS, 91405 Orsay, France}
\address{$^3$Department of Physics and Astronomy, McMaster University, Hamilton, ON, L8S 4M1 Canada}
\ead{iaffleck@physics.ubc.ca,laflorencie@lps.u-psud.fr,sorensen@mcmaster.ca}
\date{today}
\begin{abstract}
We review research on a number of situations where a quantum impurity or a physical boundary 
has an interesting effect on entanglement entropy. Our focus is mainly on impurity entanglement
as it occurs in one dimensional systems with a single impurity or a boundary, in particular quantum spin models,
but generalizations to higher dimensions are also reviewed. Recent advances in the understanding of
impurity entanglement as it occurs in the spin-boson and Kondo impurity models are discussed along with the
influence of boundaries. Particular attention is paid to $1+1$ dimensional models where analytical
results can be obtained for the case of conformally invariant boundary conditions and a connection
to topological entanglement entropy is made. New results for the entanglement in systems with mixed boundary
conditions are presented. Analytical results for the entanglement entropy obtained
from Fermi liquid theory are also discussed as well as several different
recent definitions of the impurity contribution to the entanglement entropy.

\end{abstract}
 \maketitle
\section{Introduction}
The definition of entanglement entropy is based on dividing space into two regions.  In many 
cases the systems under study are homogenuous and this division is purely fictitious. However, 
there has also been considerable activity on studying entangelement in inhomogeneous 
systems, the simplest of which contain physical boundaries or a single impurity.  
There are currently several useful measures of entanglement, here we shall use the
von Neumann entanglement entropy as defined by dividing a bipartite system in a pure
state at $T=0$ into 2 regions, $A$ and $B$.  From the ground state pure density matrix, region $B$ is traced over
to define the reduced density matrix $\rho_A$. In most cases we shall take $A$ to include the impurity/boundary. 
From this the von Neumann
entanglement entropy~\cite{Neumann27,Wehrl78},
             \begin{equation}
             S(r,R)\equiv -\Tr [\rho_A\ln \rho_A]\label{eq:vNS}
             \end{equation}
             is obtained for a subsystem of size $r$ inside a larger system of size $R$.
There are several motivations for this work.  

One motivation is to study  models of a qubit 
interacting with a decohering enviroment~\cite{Cho06,Kopp07b,Hur07c,Hur07a}. Such a system is often represented by 
a 2-level system, or spin-1/2, interacting with an otherwise homogeneous, and 
often one-dimensional medium with gapless excitations. Some versions of this 
model are equivalent to the Kondo model, motivating studies of ground state 
entanglement of an impurity spin with the conduction electrons in Kondo models. 
This entanglement 
entropy can be easily expressed exactly in terms of the impurity magnetization, 
which, for many models, has been well-understood many years ago~\cite{Hur07c,Sorensen07b}. 
This single site impurity entanglement, which we denote by $s_{\mathrm{imp}}$, is reviewed in section~\ref{sec:simp}.

Another motivation comes from the thermodynamic impurity entropy, $\ln g$, which 
was calculated by Bethe ansatz~\cite{Andrei84,Wiegmann85} for multi-channel Kondo models and then 
discussed more generally from the viewpoint of Conformal Field Theory (CFT)~\cite{Affleckg}.
 General 
quantum impurity models, such as occur in condensed matter physics, were argued to 
renormalize to conformally invariant boundary conditions, and $\ln g$ was 
argued to be a universal quantity depending only on the boundary condition.  
Calabrese and Cardy (C \& C)~\cite{Cardy04} argued that, for a CFT defined on the semi-infinite line with a conformally 
invariant boundary condition (CIBC) at the end, the entanglement of a region of length $r$ 
with the rest is given by~\cite{Cardy88,Holzhey94,Korepin04,Cardy04}:
\be S(r)=\frac{c}{6}\ln \left(\frac{r}{a}\right)+\ln g+s_1/2.\label{g}\ee
 $\ln g$ depends on the CIBC, establishing a surprising connection between 
thermodynamic and entanglement entropy.  Here $a$ is a cut-off length scale 
and $s_1$ is a
non-universal number. Both are independent of the CIBC. In the numerical work one-dimensional 
tight binding models are generally considered 
in which case $a$ is the lattice constant. This raised the possibility that this 
impurity part of the entanglement entropy, $S_{\mathrm{imp}}$, might exhibit universal renormalization 
group (RG) behavior and this was confirmed in studies of spin chains with a boundary 
magnetic field~\cite{Zhou06} and of the Kondo model~\cite{Sorensen07a,Sorensen07b}. The connection between the thermodynamic impurity entropy
and the impurity entanglement entropy
is outlined in section~\ref{sec:therment}.

There is a deep connection between (1+1) dimensional CFT and topological phases of 
gapped (2+1) dimensional systems such as occur in the fractional quantum Hall effect. 
While the entanglement entropy of a region in a gapped 2 dimensional (2D) system 
is expected to grow with the length of its perimeter, it was shown that there 
is an additional universal, length independent `` topological entropy''~\cite{Kitaev06,Levin06}, $-\ln {\cal D}$ which can 
be extracted from the corresponding (1+1) dimensional CFT. The connection 
between the boundary entropy of C \& C~\cite{Cardy04} and the topological entropy can be clarified 
by considering a (2+1) dimensional system with boundaries, a ``Hall bar'', containing a point contact. 
There are gapless degrees of freedom living on the edge of the Hall bar described by a CFT. 
(See Fig.~\ref{fig:Hall}.) 
The point contact may renormalize to a CIBC, corresponding 
to breaking the Hall bar into two pieces and the change in topological entropy 
due to this renormalization can be related to the change in the boundary 
entropy, $\ln g$.  As we will show, this is defined in terms of 
the entanglement of a section of the Hall bar of length $r$, containing the 
point contact with the rest of the (infinite) Hall bar. The connection with topological
phases is discussed in section~\ref{sec:topent}.

An alternative type of impurity-related entanglement entropy has also been studied 
for a 1D wire with a point defect, which if relevant, effective breaks 
the system in two at low energies~\cite{Levine04,Peschel05,Zhao06}. Rather than studying the entanglement 
of a finite region surrounding the point contact and extracting the $\ln g$ term 
in Eq. (\ref{g}) instead the entanglement of one side of the point contact 
with the other was studied. In cases where the point defect is relevant, 
it was found that this entanglement tends to vanish with increasing 
system size, again verifying that entanglement entropy exhibits RG flow behavior. 
This type of impurity entanglement is discussed in section~\ref{sec:peschel}.

A surprising result of numerical studies of entanglement entropy in 1D antiferromagnets with 
boundaries was the presence of an alternating term decaying away from the boundaries~\cite{Laflorencie06,Sorensen07b}.
Although 
a theory of this is still lacking, it was shown numerically to track closely 
the energy density as a function of distance from the boundary and a heuristic 
understanding was obtained in terms of a local dimerization induced by the boundary, 
related to ``resonating valence bonds''. This boundary induced alternation in the
entanglement entropy is reviewd in section~\ref{sec:altent}.

Although probably less useful as a model of a decohering environment 
and not related to CFT and universal RG concepts, impurity entanglement entropy has also been studied for 
gapped (1+1) dimensional systems, including dimerized and Haldane gap spin chains. This aspect is
reviewed in section~\ref{sec:gap}.

As a final motivation, we note that recent models for qubit teleportation and quantum state transfer
using quantum spin chains~\cite{Bose03,Christandl04,Christandl05,Burgarth05a,Burgarth05b,Wojcik05,Plenio05,Venuti06,Venuti07} employ
models closely related to the quantum spin models reviewed and in many cases rely on properties of the entanglement
arising from the impurities reviewed here.

\section{The single site impurity entanglement entropy, $s_{\mathrm{imp}}$}~\label{sec:simp}

The simplest definition of the impurity entanglement is to consider the (single site) impurity 
as sub-system $A$ and the rest of the system as sub-system $B$. A measure of the impurity
entanglement is then simply given by the von Neumann entanglement entropy of the reduced density
matrix for $A$ inside a system of total size $R$. Since $A$ describes just a single site one often refers to this as the single 
site impurity entanglement. We shall denote this quantity by $s_{\mathrm{imp}}$ to distinguish it
from $S_{\mathrm{imp}}(r,R)$ defined in later sections by the {\it difference} in the {\it uniform} part
of the von Neumann entanglement entropy for a sub-system of extent $r$ with and without the impurity present.
Since $s_{\mathrm{imp}}$ is concerned with a single site such a definition through a subtraction is not possible and
the explicit $r$ dependence through the size of the sub-system $A$ is absent.

The single site impurity entanglement, $s_{\mathrm{imp}}$, has been studied mainly in 2 different settings:
The spin-boson model~\cite{Costi03,Kopp07b,Hur07c,Hur07a} and the closely related Kondo model~\cite{Sorensen07b}.
For the case where the impurity is a $s=1/2$ system (qubit) it is easy to see~\cite{Costi03,Kopp07b,Sorensen07b} that: 
\begin{equation}
s_{\mathrm{imp}} = -\sum_{\pm}(1/2\pm m_{\mathrm{imp}})\ln [(1/2\pm m_{\mathrm{imp}})],
\label{eq:single}
\end{equation}
$m_{\mathrm{imp}}$ being the magnetization of the impurity in the ground-state.
For a system with a singlet ground state $m_{\mathrm{imp}}=0$ and $s_{\mathrm{imp}}$ is maximal~\cite{Cho06},
the qubit is maximally entangled with the rest of the system. For a system with a doublet ground-state ($R$ odd)
the behavior of $m_{\mathrm{imp}}$ is more interesting and exhibits the usual cross over associated with
Kondo physics. In this case
$m_{imp}$ was studied, for the usual fermion Kondo model, in
\cite{Sorensen96} and \cite{Barzykin96,Barzykin98} for example from which $s_{\mathrm{imp}}$ can be derived.

The spin-boson model is defined by:
\begin{equation}
H_{SB}=-\frac{\Delta}{2}\sigma_x+\frac{h}{2}\sigma_z+H_{osc}+\frac{1}{2}\sigma_z\sum_q\lambda_q(a_q+a_q^\dagger),
\end{equation}
where $\sigma_x$ and $\sigma_z$ are Pauli matrices and $\Delta$ is the tunneling amplitude between
the states with $\sigma_z=\pm 1$. $H_{osc}$ is the Hamiltonian of an infinite number of harmonic oscillators with frequencies
$\{\omega_q\}$, which couple to the spin degree via $\{\lambda_q\}$. The heat bath is characterized by its spectral function
$J(\omega)\equiv\pi\sum_q\lambda_q^2\delta(\omega_q-\omega)=2\pi\alpha\omega, \ \omega\ll\omega_c$ (Ohmic heat bath). 
Efficient NRG calculations can be performed on this model through a mapping~\cite{Anderson70} to the anisotropic Kondo model.
This allowed for rather detailed NRG studies~\cite{Costi03} of $s_{\mathrm{imp}}$ as a function of $\alpha$.
Exploiting known exact results for $\langle\sigma_z\rangle$ and $\langle\sigma_x\rangle$ in the spin-boson problem
it has been shown~\cite{Kopp07b} that $s_{\mathrm{imp}}$ in several limits is a {\it universal} function of $h/T_K$, where $T_K$ is
the Kondo scale. For $T_K\ll h\ll \Delta$ it was found~\cite{Kopp07b} that:
\begin{equation}
\lim_{T_K\ll h \ll \Delta}s_{\mathrm{imp}}(\alpha,\Delta,h)=
k_2(\alpha)\left(\frac{T_K}{h}\right)^{2-2\alpha}\ln\left(\frac{h}{T_K}\right),
\end{equation}
with $k_2(\alpha)$ a known cut-off independent funtion so that $s_{\mathrm{imp}}$ in this limit is a universal function of $h/T_K$.
On the other hand, for $h\ll T_K\ll \Delta$ it can be shown that:
\begin{equation}
\lim_{h\ll T_K \ll \Delta}s_{\mathrm{imp}}(\alpha,\Delta,h)=s_{\mathrm{imp}}(\alpha,\Delta,0)-k_1(\alpha)\left(\frac{h}{T_K}\right)^{2},
\end{equation}
with $k_1(\alpha)$ a known cut-off independent function. However, in this case $s_{\mathrm{imp}}(\alpha,\Delta,0)$ is in general non-universal
and only the second term exhibits scaling. 
\begin{figure}[!ht]
\begin{center}
\includegraphics[height=7cm,clip]{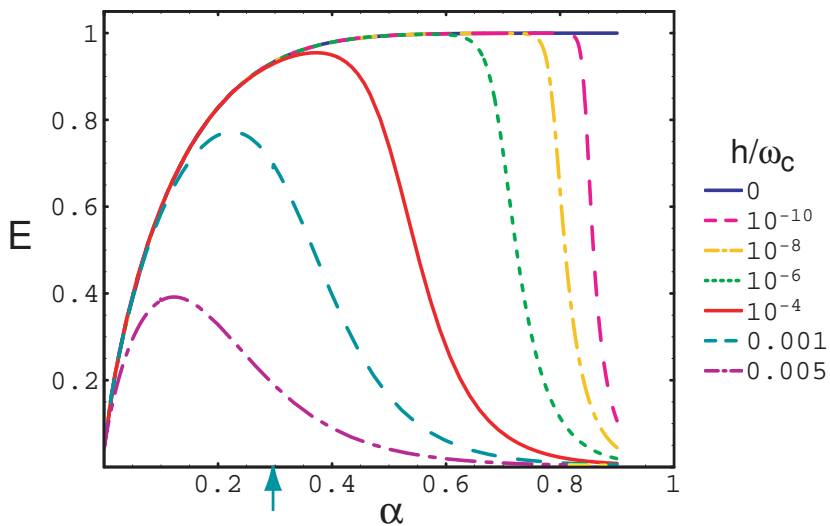}
\end{center}
\caption{
      $s_{\mathrm{imp}}(\alpha,\Delta=0.01\omega_c,h)$ ($E$ on the figure label) as a function of $\alpha$ for a range of $h/\omega_c$.
              Reproduced with permission from~\cite{Kopp07b}.
}
\label{fig:Evsalpha}
\end{figure}
The general results are illustrated in Fig.~\ref{fig:Evsalpha} where $s_{\mathrm{imp}}(\alpha,\Delta=0.01\omega_c,h)$ ($E$ on the figure label)
is plotted versus $\alpha$ for a range of $h/\omega_c$. For $h=0$ $s_{\mathrm{imp}}(\alpha)$  is a monotonically increasing function where as
for non-zero $h$ a maximum associated with the crossover $h\sim T_K$ occurs.
The case of a sub-Ohmic heat bath has also been studied~\cite{Hur07c}.

The usual Kondo Hamiltonian~\cite{KondoRMP,Hewson} contains a Heisenberg interaction between a 
$s=1/2$ impurity spin, $S$,
and  otherwise non-interacting electrons.
A simple model takes a free electron dispersion relation and a
$\delta$ -function Kondo interaction:
\begin{equation}
H=\int d^{3}r[\psi ^{\dagger }(-\nabla ^{2}/2m)\psi +J _{K}\delta^3(\vec
    r)\psi ^{\dagger }(\vec{\sigma}/2)\psi \cdot \vec{S}].  \label{H3DKondo}
\end{equation}
(Actually, an ultra-violet cut-off of the $\delta$-function interaction is necessary for 
the model to be completely well-defined.)
In the ground-state of the Kondo model the impurity spin is screened by the conduction electrons through the formation
of a singlet. This phenomenon is expected to take place on a length scale:
\begin{equation}
\xi_K=v/T_K\propto e^{1/(\nu J_K)},
\end{equation}
where $\nu$ is the density of states per spin band , $T_K$ is the Kondo scale and $v$ the velocity of the fermions. 
Due to the $\delta-$ function form of the interaction
Eq.~(\ref{H3DKondo}) can be reduced to a one-dimensional model which can be represented by a lattice model of finite extent $R$ (including the impurity), suitable
for numerical studies. As outlined above and discussed in more detail in section~\ref{sec:boundent}, we expect the case of $R$ {\it odd} to reflect Kondo physics and possibly scaling with $R/\xi_K$.
However, it has been shown~\cite{Sorensen07b} 
that $s_{\mathrm{imp}}$ exhibits weak scaling violations from the expected $R/\xi_K$ scaling. In particular, $m_{\mathrm{imp}}$ was
shown to take the form:
\begin{equation}
m_\mathrm{imp}\sim \frac{1}{2}-\left[\frac{J_K}{\pi v}\right]^2\ln(R/a),
\end{equation}
where $a$ is a short distance cut off. Hence weak scaling violations are present in $m_{\mathrm{imp}}$ and therefore also in $s_{\mathrm{imp}}$
even though $s_{\mathrm{imp}}$ clearly displays the expected crossover related to Kondo physics.
\begin{figure}[!ht]
\begin{center}
\includegraphics[height=7cm,clip]{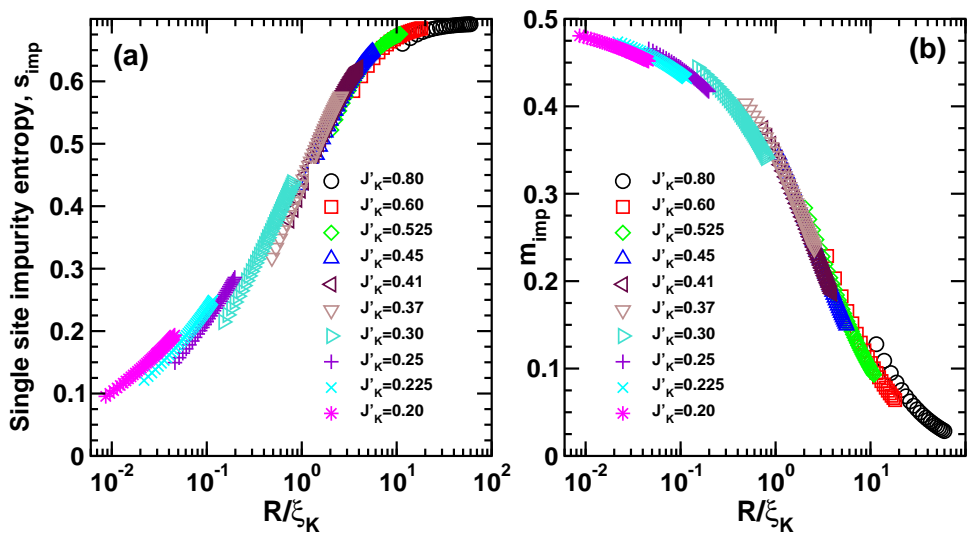}
\end{center}
\caption{
  (a) Weak scaling violations for the single site impurity entanglement entropy $s_{imp}$ (a) [Eq.~(\ref{eq:single})]
    and the local impurity magnetization $m_{imp}=\langle S_{1}^{z}\rangle$ (b).
    All data are for odd length chains between $R=19\ldots 101$. Reprinted from~\cite{Sorensen07b}.
}
\label{fig:violation1}
\end{figure}
The presence of such scaling violations is illustrated in Fig.~\ref{fig:violation1}.

Finally, recent work has focused on the entanglement between the 2 impurity spins in the two-impurity Kondo model~\cite{Costa06,Cho06, Samuelson07} 
as well as the pair-wise entanglement between end spins in open ended Heisenberg spin chains~\cite{Hu08}.

\section{Impurity entropies from the CFT perspective}\label{sec:therment}
In this section we review the concept of thermodynamic impurity entropy and its connection with 
CFT. We also connect it with topological entropy in 2D topological insulators and 
discuss various applications of these general ideas to specific models: spin chains with boundary fields, Kondo model and 
point contact in a Hall bar. 
\subsection{Thermodynamic Impurity Entropy}\label{sec:2.1}
Thermodynamic impurity entropy can be measured experimentally from impurity specific heat using the 
thermodynamic identity:
\be C=T{\partial S\over \partial T}.\ee
The impurity contribution to the specific heat, and hence the entropy, can be measured by subtracting 
off the same quantity for the pure system. Such measurements and related theoretical calculations
have been performed for many years in Kondo impurity systems. 

$C(T)$ for a metal with a dilute random array of magnetic impurities is measured 
and $C(T)$ for the pure system is subtracted off.  Dividing by the number of impurities 
and extrapolating to zero impurity density gives the contribution to the specific 
heat of a single impurity, $C_{\mathrm{imp}}(T)$.  From integrating $C_{\mathrm{imp}}(T)/T$ 
one can, at least in principle, obtain $S^{\mathrm{Th}}_{\mathrm{imp}}(T)$.  We consider the 
infinite volume limit, so that $S^{\mathrm{Th}}_{\mathrm{imp}}(T)$ becomes a function of $T$ only. 
We may formally define $S^{\mathrm{Th}}_{\mathrm{imp}}(T)$ by calculations on a 1D system of length $R$ with a 
single impurity:
\be S^{\mathrm{Th}}_{\mathrm{imp}}(T)\equiv \lim_{R\to \infty}[S(T,R)-S_0(T,R)].\ee
Here $S(T,R)$ is the thermodynamic entropy with the impurity present and 
$S_0(T,R)$ is the same quantity without the impurity. $S^{\mathrm{Th}}_{\mathrm{imp}}(T)$ exhibits 
interesting $T$-dependence which reflects the RG flow of the Kondo model. It can 
be calculated with high precision from the Bethe ansatz solution of the Kondo 
model first derived by Andrei~\cite{Andrei80} and Wiegmann~\cite{Wiegmann80}.  In the limit of a weak bare Kondo 
coupling $S^{\mathrm{Th}}_{\mathrm{imp}}(T)\approx \ln 2$ at $T\gg T_K$ and
$S^{\mathrm{Th}}_{\mathrm{imp}}(T)\to 0$ at $T\ll T_K$. This reflects the fact that the 
bare coupling of the magnetic impurity to the conduction electrons is very weak 
so that we obtain essentially the full entropy of a free spin-1/2, $\ln 2$ at 
$T\gg T_K$. However, as the temperature is lowered the spin is ``screened'' i.e. 
it goes into a singlet state and the impurity entropy is accordingly lost. The 
asymptotic values of $S^{\mathrm{Th}}_{\mathrm{imp}}(T)$ at high and low temperatures are characteristic 
of the RG fixed points of the Kondo Hamiltonian. 

The $k$-channel Kondo model 
\begin{equation}
H=\int d^{3}r\sum_{i=1}^k[\psi ^{\dagger i }(-\nabla ^{2}/2m)\psi_i +J _{K}\delta^3(\vec
    r)\psi ^{\dagger i}(\vec{\sigma}/2)\psi_i \cdot \vec{S}].  \label{kH3DKondo}
\end{equation}
also has $S^{\mathrm{Th}}_{\mathrm{imp}}(T)\approx \ln 2$ at $T\gg T_K$.  However, it was found, 
from the Bethe ansatz solution, that at $T\to 0$ it has the limiting value $\ln g$ 
where~\cite{Affleckg}:
\be g=2\cos [\pi /(2+k)]\leq 2.\label{gKondo}\ee
Heuristically, $g$ represents a ``fractional ground state degeneracy'' characterizing 
the non-Fermi liquid ground state of the overscreened Kondo model. 

The interesting behavior of the impurity entropy in the multi-channel Kondo 
model was later shown to be a special case of a general phenomenon in quantum 
impurity models. Many models of this type have low energy descriptions 
in terms of 1D CFT's. Quite general boundary conditions and boundary interactions 
are expected to renormalize, at low energies, to CIBC's. Cardy showed 
that generally  CIBC's can be associated with boundary states, $|A>$. The 
conformally invariant partition function defined on a strip of length $R$, 
at inverse temperature $\beta$, with CIBC's $A$ and $B$ at the two ends 
can be written:
\be Z_{AB}=\mathrm{tr}e^{-\beta
H^R_{AB}},\label{ZAB1}\ee
where $H^R_{AB}$ is the Hamiltonian on an interval of length $R$ with 
BC's $A$ and $B$ at the two ends.  Alternatively, we may switch space 
and imaginary time directions and write:
\be  Z_{AB} =
<A|e^{-RH^\beta_P}|B>.\label{ZAB2}\ee
Now the system propagates for time $R$, under the action of the Hamiltonian 
defined on a periodic interval of length $\beta$ with initial and final 
states $|A>$ and $|B>$. The boundary states can be expanded in a complete 
basis of Ishibashi states, associated with each conformal tower, $a$:
\be |A> = \sum_a|a><a0|A>.\ee
(The Ishibashi states $|a>$ are sums over all descendants with equal 
weight in left and right-moving sectors.) Thus, we may write:
\begin{eqnarray} Z_{AB} &=&
\sum_a<A|a0><a0|B><a|e^{-RH^\beta_P}|a>\nonumber\\
&=&\sum_a<A|a0><a0|B>\chi_a (e^{-4\pi R/\beta})
\end{eqnarray}
where $\chi_a$ is the character of the $a^{\mathrm{th}}$ conformal tower.
Written in this form it is straightforward to extract the impurity entropy.  
Taking the limit $R\gg \beta$ only the highest weight state in the conformal tower of the identity operator 
contributes, giving:
\be Z_{AB}\to  e^{\pi Rc/6\beta}<A|00><00|B>.\ee
(Here $c$ is the central charge associated with the bulk CFT.) 
From this expression we obtain the entropy:
\be S^{\mathrm{Th}}_{AB}={\partial \over \partial T}[T\ln Z_{AB}] ={\pi RcT\over 3}+\ln g_{AB}\ee
where:
\be g_{AB}=g_A\cdot g_B= <A|00><00|B>.\ee
This consists of the bulk term, independent of the boundary conditions and proportional 
to the system size, $R$, as well as the boundary term $\ln g$ which is a sum of 
contributions from each boundary.  Using the known boundary state corresponding 
to the non-Fermi liquid ground state of the multi-channel Kondo model we 
can reobtain the Bethe ansatz result for $S^{\mathrm{Th}}_{\mathrm{imp}}(T=0)$ from this general CFT formula. 

$\ln g$, the $T=0$ thermodymamic entropy, is a universal property of fixed points 
of the boundary RG. It has the interesting property that it always {\it decreases} 
under RG flow from an unstable to stable fixed point~\cite{Affleckg}. Eq. (\ref{gKondo}) 
provides an example of this: the impurity entropy is $\ln 2$ at the unstable 
fixed point, $T\gg T_K$, and always has a smaller value at the stable fixed point which 
occurs at $T=0$.

\subsection{Boundary term in the entanglement entropy}\label{sec:boundent}
We now consider a semi-infinite CFT ($r\geq 0$) with CIBC, of type $A$ at $r=0$. We 
consider the ground state entanglement entropy for the region, $0\leq r' \leq r$, $S_A(r)$. 
We might now expect some additional term in $S(r)$, $c_A$, which depends on the CIBC $A$, 
but not on the length of the region under consideration, $r$:
\be S_A(r)=\frac{c}{6}\ln \left(\frac{r}{a}\right)+c_A+\frac{s_1}{2}.\ee
We can argue that $c_A=\ln g_A$, the thermodynamic impurity entropy, by the device 
of considering the entanglement entropy for this system at finite temperature. 
C\&C~\cite{Cardy04} showed that the generalization of $S_A(r)$ to a finite inverse temperature, $\beta$, is given by a 
standard conformal transformation:
\be S_A(r,\beta )=\frac{c}{6}\ln \left[\frac{\beta }{\pi a}\sinh \left(\frac{2\pi r}{\beta }\right)\right]+c_A+s_1/2.\ee
See also \cite{Korepin04}.
$S_A(r,\beta )$ is defined by beginning with the Gibbs density matrix for the entire system, $e^{-\beta H}$ 
and then again tracing out the region $r'>r$. 
Now consider the high temperature, long length limit, $\beta \ll r$:
\be S_A\to \frac{2\pi cr}{6\beta} +\frac{c}{6}\ln \left(\frac{\beta }{2\pi a}\right)+c_A+\frac{s_1}{2}+O(e^{-4\pi r/\beta}).\label{Sas}\ee
The first term is the extensive term (proportional to $r$) in the thermodynamic entropy for 
the region, $0<r'<r$. 
The reason that we recover the thermodynamic entropy when $r\gg \beta$ is because, 
in this limit, we may regard the region $r'>r$ as an ``additional reservoir'' for the 
region $0\leq r'\leq r$. That is, the thermal density matrix can be defined by integrating 
out degrees of freedom in a thermal reservoir, which is weakly coupled to the entire system. 
On the other hand, the region $r'>r$ is quite strongly coupled to the region $r'<r$. 
Although this coupling is quite strong, it only occurs at one point, $r$. When 
$r\gg \beta$, this coupling only weakly perturbs the density matrix for the region $r'<r$. 
Only low energy states, with energies of order $1/r$ and a neglegible fraction of the 
higher energy states (those localized near $r'=r$) are affected by the coupling 
to the region $r'>r$. The thermal entropy for the system, with the boundary at $r=0$ in the limit $r\gg \beta$ is:
\be S^{\mathrm{Th}}_{A}\to \frac{2\pi cr}{6\beta} + \ln g_A+\mathrm{constant},  \ee
with corrections that are exponentially small in $r/\beta$.
The only dependence on the CIBC, in this limit, is through the constant term, $\ln g_A$, the impurity entropy. 
Thus it is natural to identify the BC dependent term in the entanglement entropy 
with the BC dependent term in the thermodynamic entropy:
\be c_A=\ln g_A.\ee
This follows since, in the limit, $r\gg \beta$, we don't expect the coupling to the 
region $r'>r$ to affect the entanglement entropy associated with the boundary $r'=0$, $c_A$. 
Note that the entanglement entropy, Eq. (\ref{Sas}), contains an additional large term 
not present in the thermal entropy. We may ascribe this term to a residual effect of the 
strong coupling to the region $r'>r$ on the reduced density matrix. However, this extra term does not depend on the CIBC 
as we would expect in the limit $r\gg\beta$ in which the ``additional reservoir'' is far from the boundary. 
Now passing to the opposite limit $\beta \to \infty$, we obtain the remarkable result that 
the only term in the (zero temperature) entanglement entropy depending on the BC is precisely 
the impurity entropy, $\ln g_A$, Eq. (\ref{g}).  We may also consider a finite system, 
of length $R$, with ICBC at both ends. In the case, where both BC's are the same, $A$, 
we expect the generalization of Eq. (\ref{g}) which follows from a conformal transformation:
\be S_A(r,R)=\frac{c}{6}\ln \left[\frac{2R}{\pi a}\sin \left(\frac{\pi r}{R}\right)\right]+
\ln g_A+\frac{s_1}{2},\label{gL}\ee
with the equivalent result for periodic boundary conditions:
\be S^{\mathrm{PBC}}_A(r,R)=\frac{c}{3}\ln \left[\frac{R}{\pi a}\sin \left(\frac{\pi r}{R}\right)\right]+s_1.\label{Spbc}\ee
As usual, $a$ denotes the lattice spacing.
A result has not been obtained, as far as we know, for the case where the CIBC's 
are different at the two ends.

As discussed in the previous sub-section, the thermodynamic impurity entropy, $\ln g$, 
is a universal quantity with interesting RG behavior. It is then natural to expect 
that the impurity term in the entanglement entropy will behave similarly. In particular, 
consider beginning with a CFT on the semi-infinite line with CIBC $A$ and then adding 
a small relevant boundary interaction:
\be H=H_{\mathrm{CFT}}-\lambda \phi (0) \ee
where $\phi$ has an RG scaling dimension $y<1$. ($y=1$ is the marginal dimension 
for {\it boundary} interactions, since the action contains only a time-integral 
over $\phi (\tau ,r=0)$ and no spatial integral.)  We expect that, under the renormalization 
group, the Hamiltonian will flow to an infrared stable fixed point, at long length 
scales, characterized by some other CIBC, $B$.  The ``g-theorem''~\cite{Affleckg,Friedan04} implies 
that the thermodynamic impurity entropy at the stable fixed point obeys $g_B<g_A$. 
Typically the flow between fixed points can be controlled by introducing a finite length 
scale which acts as an infrared cut off. This is often done by putting the system 
in a finite box, of size $R$. It is not obvious what will happen when we have 
no physical box but introduce a length scale $l$ by the definition of the 
entanglement entropy. Will the impurity part of the entanglement entropy exhibit 
a cross over from $g_A$ to $g_B$ as we increase $l$? Will this cross over be 
universal? These questions have been investigated numerically, using the DMRG 
method, in a couple of models. 
\begin{figure}[!th]
\begin{center}
\includegraphics[width=0.8\columnwidth,clip]{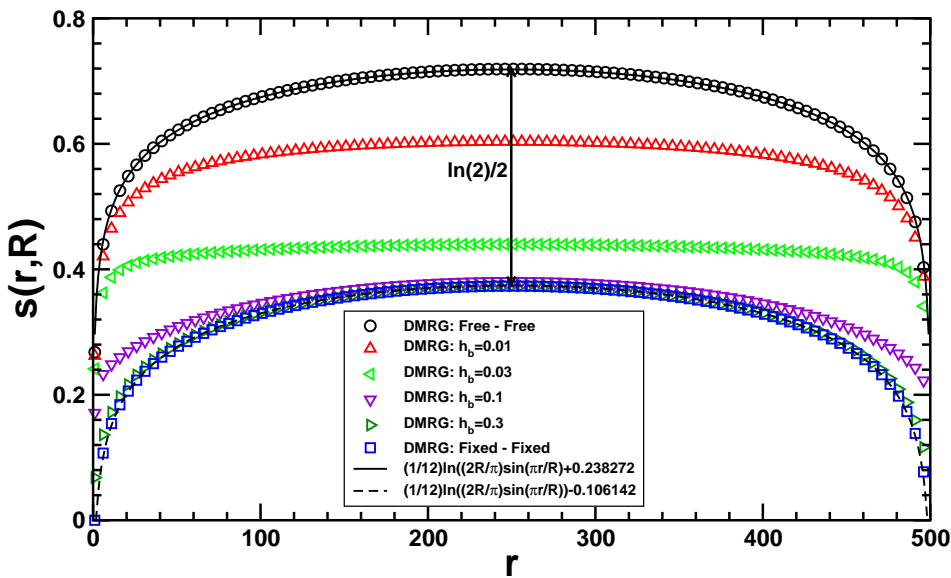}
\end{center}
\caption{
  $S(r,R)$ versus $r$ ($R=500$) for the transverse field Ising model calculated using DMRG with $m=128$. Results are
  shown for free-free and fixed-fixed boundary conditions along with 
  four different values of $h_b=0.01,0.03,0.10,0.30$. For the numerical DMRG data only every 4th point is shown for
  clarity. The solid and dashed lines represent fits of the DMRG results for free-free and fixed-fixed BC's, respectively,
  to the analytical result Eq.~(\ref{gL}). For the latter case calculations were performed formally with $h_b=\infty$.
} \label{fig:hbRL}
\end{figure}

\begin{figure}[!th]
\begin{center}
\includegraphics[width=0.8\columnwidth,clip]{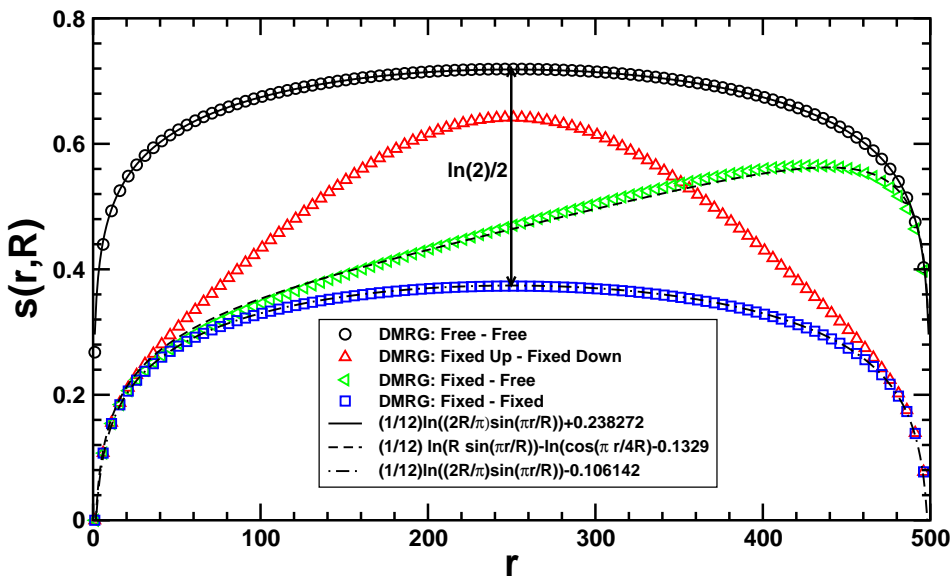}
\end{center}
\caption{
  $S(r,R)$ versus $r$ ($R=500$) for the transverse field Ising model calculated using DMRG with $m=128$. Results are
  shown for free-free and fixed-fixed boundary conditions along with free-fixed 
  and fixed (up) - fixed (down) boundary conditions. For the numerical DMRG data only every 4th point
  is shown for clarity. The solid and dashed lines represents fits of the DMRG data for free-free and fixed-fixed
  BC's, respectively, to the analytical result Eq.~(\ref{gL}). The calculations with fixed boundaries are performed
  by formally setting $h_b=\pm \infty$.
} \label{fig:AllBC}
\end{figure}

The first model considered was the 1D transverse field Ising model with a 
longitudinal boundary magnetic field~\cite{Zhou06}:
\be H=-\sum_{j=0}^\infty (S^z_jS^z_{j+1}+(1/2)S^x_j)+h_bS^z_0.\ee
The bulk transverse field has been tuned to its critical value, $1/2$. 
A weak longitudinal  field, $h_b$ is applied at the boundary, $j=0$, only. 
This boundary field is relevant, with dimension 1/2 and therefore induces a boundary RG flow between the only  two 
boundary fixed points in this model, corresponding to free or fixed BC.  
The values of $g$ are $g=1$ (free) and $g=1/\sqrt{2}$ (fixed). 
DMRG results on chains of length up to 800, keeping 140 block states, 
showed quite convincingly that the entanglement entropy crosses over from 
$(1/12)\ln (r/a)$ to $(1/12)\ln (r/a)-(1/2)\ln 2$ as $r$ is increased from 
small values to values larger than a cross-over scale, $\xi$, determined by $h_b$ as
$\xi\propto |h_b|^{-1/2}$.  In figure~\ref{fig:hbRL} we illustrate these results by calculations
on systems with $R=500$.
In figure~\ref{fig:AllBC} we also show the entanglement entropy with {\it different} BC's at the 
ends of a finite chain, fixed-free and fixed up-fixed down. As far as we know, no 
analytic formulas have been derived for these cases. We note that the expression:
\begin{equation}
S(r,R)={1\over 12}\ln \left[{R\over a}\sin \left({\pi r\over R}\right) \right]-
\ln \left[\cos \left({\pi r\over 4R}\right)\right] 
+\hbox{constant}\end{equation}
seems to fit the data quite well in the fixed-free case. 

The second model in which this crossover was studied is the Kondo model~\cite{Sorensen07a,Sorensen07b}
 For this model, we may consider 
the entanglement of a region inside a sphere of size $r$ surrounding the impurity, with the rest of space 
which may either be infinite or confined to a larger sphere of size $R$. The 
impurity entanglement entropy is defined as the increase in entanglement arising 
when the impurity is added to the system.  This mimics the definition of the 
impurity thermodynamic entropy which has been well-studied experimentally 
and theoretically for Kondo systems. Due the $\delta$-function form of the Kondo-interaction, 
the 3D model is equivalent to a 1D model.  To see this, we expand the electron fields in 
spherical harmonics.  Only the s-wave interacts with the impurity spin, the other 
harmonics being completely free. The entanglement entropy may be written as a sum of 
contributions from each spherical harmonic:
\be S=\sum_{l,m}S_{l,m}.\ee
Only the s-wave part, $S_{0,0}$ is affected by the Kondo interaction so that the total 
impurity entanglement entropy is given by the change in $S_{0,0}$ when the impurity spin is added. 
Assuming that the Kondo coupling is weak, as is usually the case in experiments, we may 
integrate out Fourier components of the s-wave electron fields except for a narrow band 
around the Fermi sphere. Linearizing the dispersion relation near the Fermi energy:
\be k^2/2m\approx k_F^2/2m + v_F(k-k_F),\ee
the model becomes equivalent to a relativistic Dirac fermion defined on the half-line, $r>0$ 
interacting with the impurity spin at $r=0$:
\begin{eqnarray}H_{1D} \approx&& (iv_F/2\pi )\int_{0}^{R}dr[\psi _{L}^{\dagger
}(d/dr)\psi _{L}-\psi _{R}^{\dagger }(d/dr)\psi _{R}]  \nonumber\\
&&+v_F\lambda_{K}\psi _{L}^{\dagger }(0)(\vec{\sigma}/2)\psi _{L}(0)\cdot \vec{S}.
\label{H1D}
\end{eqnarray}
Here $\lambda_K\propto J$ and a boundary condition:
\be \psi_L(0)=-\psi_R(0)\label{freebc}\ee
is imposed on the left and right movers. 

To obtain a model amenable to DMRG studies, 
we could consider a 1D tight-binding version of this 1D continuum model. However, considerable 
numerical speed-up can be obtained by considering a ``spin-only'' version of the model. 
This is based on spin-charge separation for 1D interacting fermion systems, which 
follows from bosonization. We find that only the spin degrees of freedom of the 
1D electrons interact with the impurity, {\it when it is at the end of the chain}.  
[Things are more complicated when it is not at the end.  The simplifications at the end 
arise from the BC of Eq. (\ref{freebc}).]
The spin part of the Hamiltonian, the only part involving the Kondo interaction, can 
be written as a perturbed  $SU(2)_1$ Wess-Zumino-Witten non-linear $\sigma$-model with 
Hamiltonian:
\be 
H_{s}=(v_F/2\pi )\int_0^{R}dr (1/3)[\vec J_L\cdot \vec J_L+\vec J_R\cdot \vec J_R]
+v_F\lambda_K\vec J_L(0)\cdot \vec S.\label{Hspin}\ee
Here $\vec J_{L/R}(r)$ are the spin density operators for left and right movers, 
with the BC
\be \vec J_L(0)=\vec J_R(0).\ee
 This implies that we may regard 
$\vec J_R(r)$ as the analytic continuation of $J_L(r)$ to the negative $r$ 
axis:
\be J_R(r)=\vec J_L(-r)\ee
and write the theory in terms of left movers only defined on the interval 
$-R<r<R$:
\be H_{s}=(v/6\pi )\int_{-R}^{R} dr \vec J_L\cdot \vec J_L+v_F\lambda_K\vec J_L(0)\cdot \vec S.
\label{HsL}\ee

  Now consider Heisenberg antiferromagnetic S=1/2 chain with one weak link at the end:
\be H=J_{K}^{\prime }\vec{S}_{1}\cdot \vec{S}_{2} +
\sum_{r=2}^{R-1}\vec{S}_{r}\cdot \vec{S}_{r+1}.\ee
For $J_{K}^{\prime }\ll 1$ essentially the same low energy continuum limit field theory, 
Eq. (\ref{Hspin}) is obtained except that the Fermi velocity, $v_F$ is replaced 
by the spin-velocity, which we call $v$. This model is considerably more efficient to study with 
DMRG since there are only 2 states per site rather than 4. Actually, a drawback of 
this model is that there is an important marginally irrelevant bulk interaction in the 
low energy Hamiltonian: 
\be \delta H = -(gv/2\pi )\vec J_L\cdot \vec J_R.\label{margint}\ee 
with the positive dimensionless coupling constant $g$ of O(1).  This leads to logarithmically 
varying corrections to all quantities which greatly hinders numerical work. 
To circumvent this problem, it is advantageous to add a second neighbor interaction, 
considering instead the Hamiltonian:
\begin{equation}
H =J_{K}^{\prime }\left( \vec{S}_{1}\cdot \vec{S}_{2}+J_{2}\vec{S}%
_{1}\cdot \vec{S}_{3}\right) +
\sum_{r=2}^{R-1}\vec{S}_{r}\cdot \vec{S}_{r+1}+J_{2}\sum_{r=2}^{R-2}\vec{%
S}_{r}\cdot \vec{S}_{r+2}.  \label{eq:spinch}
\end{equation}
For $J_2>J_2^c\approx 0.2412$ the model goes into a gapped dimerized phase, of which 
the exactly solvable Majumdar-Ghosh model with $J_2=1/2$ is a special simple case. 
The gap is driven by the marginal coupling constant $g$ which changes sign 
at $J_c^c$, becoming marginally relevant. 
For $J_2<J_2^c$ the model remains gapless with the marginal coupling constant $g$ and 
the spin velocity $v$ varying smoothly. Right at the critical point, $J_2=J_2^c$, $g=0$. 
At this point all logarithmic corrections vanish and it becomes possible to 
extract meaningful results from numerical studies of relatively short chains. 
Therefore, we largely focussed on this $J_2=J_2^c$ model.  As discussed in more detail in Ref.~\cite{Laflorencie08}
and references therein, we then see that the low energy effective field theory description of the spin only model, Eq.~(\ref{eq:spinch}),
with $J_2=J_2^c\ (g=0)$ is {\it the same} as that of the usual electronic version of the Kondo model.

As summarized in sub-section~\ref{sec:2.1} , the thermodynamic impurity entropy, $S^{\mathrm{Th}}_{\mathrm{imp}}(T)$ 
decreases monotonically from $\approx \ln 2$ at $T\gg T_K$ to zero at $T\ll T_K$. 
We might expect that the impurity entanglement entropy would behave the same 
same with the energy scale, $T$ replaced by $v/r$ where $r$ is the size of the 
region $A$. While we ultimately confirmed this result two interesting subtleties 
were encountered en route. 

First of all, even for an infinite system size, $R$, 
we found that the entanglement entropy has an alternating term, $S_A$, which decays 
only slowly with $r$~\cite{Laflorencie06}:
\begin{equation}
S(r,R)=S_U(J'_K,r,{R})+(-1)^rS_A(J'_K,r,{R}),
\end{equation}
We discuss this in section~\ref{sec:altent}. Although we expect $S_U$ to be essentially the same 
in the spin chain Kondo model as in the fermion Kondo model, the same is not 
true of $S_A$.  (More generally, the fermion model is expected to have a term 
in the entanglement entropy oscillating at wave-vector $2k_F$.) Henceforth, in this section, 
we focus on the uniform part, $S_U$ only. 

\begin{figure}[!th]
\begin{center}
\includegraphics[width=0.7\columnwidth,clip]{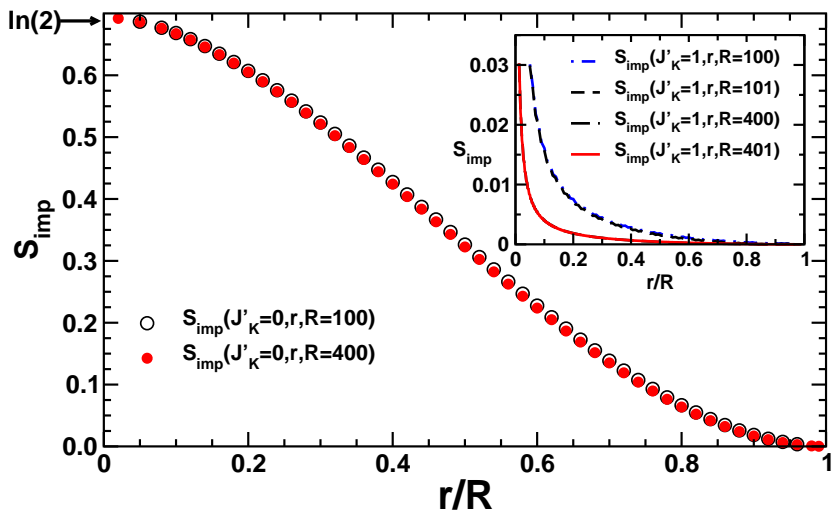}
\end{center}
\caption{
  DMRG results using spin-inversion with $m=256$for $S_{\mathrm{imp}}(0,r,{R})$  for $\mathcal{R}$ even (${R}=100,400$).
    Results are shown for the $J_1-J_2$ spin chain model at $J_2^c$.
    Inset: DMRG results for $S_{\mathrm{imp}}(1,r,{R})$  for ${R}$ odd and even. Data for ${R}=100,101$
    and ${R}=400,401$ are indistinguishable.
    $S_{{imp}}(1,r,{R})$  appears to vanish in the scaling limit $r\to\infty$ with $r/{R}$
    held fixed.
} \label{fig:sfp}
\end{figure}

\begin{figure}[th]
\begin{center}
\includegraphics[height=6cm,clip]{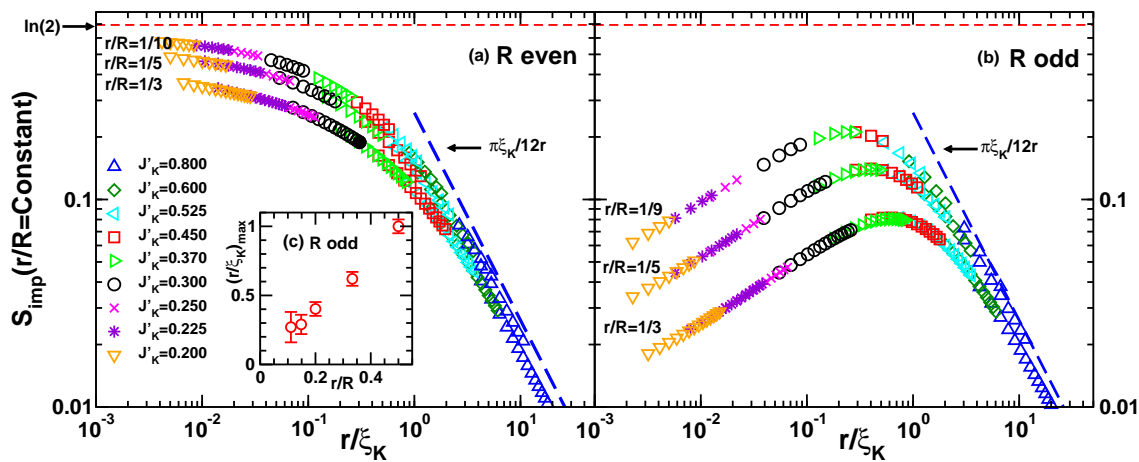}
\end{center}
\caption{Universal scaling plot of $S_{\mathrm{imp}}$ for fixed $r/R$,
  (a) for $R\leq 102 $ {\it even}, (b) for $R\leq 101 $ {\it odd}.
    DMRG results with $m=256$ for the $J_{1}-J_{2}$ chain at $J^c_2$
    for various couplings $J_{K}^{\prime }$.
      The lines marked $ \pi \xi_K/(12r)$ are the FLT
        prediction: Eq.~(\ref{eq:FLT_FSS}).
        (c): the location of the maximum, $(r/\xi_K)_{\rm max}$,
        of $S_{\mathrm{imp}}$ for odd $R$, plotted versus $r/R$.
          Reprinted from ~\cite{Sorensen07b}.
} \label{fig:kondoDMRG}
\end{figure}

We extracted an impurity part from $S_U$ by:
\begin{equation}
S_{\mathrm{imp}}(J'_K,r,{R})\equiv S_U(J'_K,r,{R})-S_{U}(1,r-1,{R}-1),\ r>1.\label{Simpdef}
  \end{equation}
Note that we are subtracting the entanglement entropy of a chain where all couplings 
have unit strength and 1 site is removed.  This corresponds to subtracting the 
entanglement entropy of the system without the impurity. It is important here 
that we do not subtract the entropy for the same values of $r$ and $R$ but with $J_K'=0$ 
since, as we discuss below, entanglement with the impurity can survive, even in this limit, 
when $R$ is even, in the spin-singlet ground state. (See Fig.~\ref{fig:sfp}). This is related to the second 
subtlety that we encountered: a very strong dependence of $S_{\mathrm{imp}}$, as defined 
by Eq. (\ref{Simpdef}), on the 
parity of $R$, even after subtracting off the part alternating in $r$. This is 
illustrated by some of our DMRG data shown in Fig. \ref{fig:kondoDMRG}.

This figure tests the conjecture that the impurity entanglement entropy shows universal 
RG scaling behavior.  We find that the data for the impurity entanglement entropy 
for various Kondo couplings can 
be collapsed onto scaling curves which depend only on the dimensionless ratios 
$r/\xi_K$, where $\xi_K$ is the Kondo length scale, and $r/R$.  However, there 
are two different sets of curves depending on the parity of $R$. Note that 
the curves differ markedly for $r\ll \xi_K$ but elsewhere look similar.  Indeed 
it looks likely, and presumably must be the case, that as $r/R\to 0$, the 
curves become identical for even and odd $R$. Focusing on the even $R$ curves, 
which seem close to the $r/R\to 0$ limit, the scaling curves seem to be 
approach a monotone decreasing function with the value $\ln 2$ at $r\ll \xi_K$ 
and zero for $r\gg \xi_K$. This is exactly what we would expect from the 
RG theory of the Kondo model and mirrors the $T$-dependence of the 
thermodynamic impurity entropy, reviewed in subsection (3.1).  In particular, 
$\ln g = \ln2$ at the short distance weak coupling fixed point, corresponding 
to a paramagnetic spin-1/2 impurity but $\ln g=0$ at the long distance 
strong coupling fixed point, corresponding to the spin being screened. 

A qualitative understanding of the surprising difference between even and odd $R$ 
can be obtained using a ``resonating valence bond'' picture of the ground 
state of the $s=1/2$ Heisenberg antiferromagnetic chain~\cite{Sorensen07b}.
Consider first even $R$. Any singlet 
state can be written as a linear superposition of products of 
singlet states formed by pairs of spins. Conventionally one draws a 
line or ``valence bond'' between pairs of spins contracted to a singlet, 
$(|\uparrow \downarrow >-|\downarrow \uparrow >)/\sqrt{2}$. It can 
easily be shown that by restricting to terms in which none of the 
lines cross we get a complete linearly independent set of singlet states 
for a S=1/2 chain. Furthermore, by adopting a convenient sign 
convention for valence bond states, it can be proven 
that all terms in the sum have non-negative coefficients. We may heuristically associate the impurity 
entanglement entropy with the valence bond originating from the 
impurity spin at site 1. If this spin forms a singlet with a spin 
at a site inside region $A$ (at a site with index $\leq r$) then 
we consider this not to contribute to the impurity entanglement 
entropy, $S_{\mathrm{imp}}$ of region $A$.  On the other hand, if site 1 is paired 
with a site outside region $A$ (with index $>r$) then we 
consider it to contribute to $S_{\mathrm{imp}}$. 
\begin{figure}[th]
\begin{center}
\includegraphics[height=3.5cm,clip]{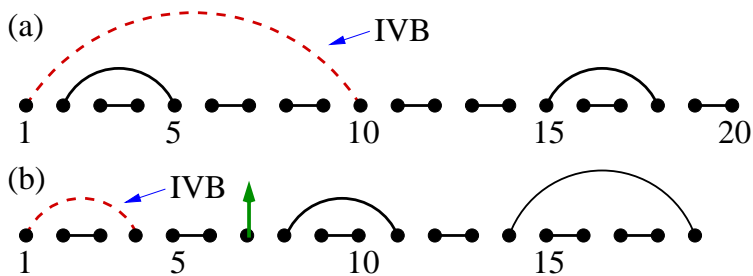}
\end{center}
\caption {
  Typical impurity  valence bond configurations for $R$ even (a) and
    $R$ odd (b). Note the unpaired spin on the $7^{\rm th}$ site in (b).
    Reprinted from~\cite{Sorensen07b}.
}
\label{fig:ivb}
\end{figure}
The ground state is 
a sum over many valence bond configurations so we could imagine 
relating $S_{\mathrm{imp}}$ to the probability of this ``impurity valence bond'' (IVB) 
extending out of region $A$~\cite{Sorensen07a,Sorensen07b}. 
(See Fig.~\ref{fig:ivb}.) Similar ideas have been
explored in~\cite{Refael04}. As $J_K'$ gets smaller, the IVB gets 
longer ranged. This follows because, in general, valence bonds 
tend to be nearest neighbor, or at least quite short range 
in order to take advantage of the nearest neighbor antiferromagnetic 
interactions.  However, as $J_K'$ gets weaker there is less 
and less energetic advantage in a short impurity valence bond. 
In the extreme case $J_K'=0$ the impurity valence bond 
extends with significant probability over the entire chain.  
We expect that the typical length of this impurity valence bond 
should be $\xi_K$, the Kondo screening cloud size. This is a characteristic 
length scale, of order $v/T_K$, at which the effective Kondo 
coupling becomes of order one. In a metal, one heuristically 
imagines an electron in a quasi-bound state forming a singlet 
with the impurity spin with $\xi_K$ being the extent of the 
bound state wave-function. In the spin chain realization of the Kondo 
model this quasi bound state corresponds to the impurity valence bond. 
Thus the fact that $S_{\mathrm{imp}}$ starts to decrease when $r$ exceeds $\xi_K$ 
is very natural from this Kondo screening cloud viewpoint. 
Unfortunately, it seems extremely difficult to make this more than 
a heuristic argument. A problem is that the valence bond basis, 
while complete and linearly independent, is not an orthonormal basis. 
Although one could define a probability distribution for 
the length of the IVB it is hard to relate this to physical quantities 
such as the correlation length or the entanglement entropy. However, we
can formally write:
\begin{equation}
S_{\mathrm{imp}}^{\mathrm{IVB}}=(1-p)\ln 2,
\end{equation}
with $p$ the probability that the IVB connects the impurity spin to a spin in region $A$. This
follows, since if the IVB does {\it not} cross the boundary between regions $A$ and $B$
its contribution to $S_{\mathrm{imp}}$ is obviously zero, on the other hand, if the IVB connects to 
a spin in region $B$ it will contribute a factor of $\ln(2)$. 
It is possible to be a little bit more quantitative using the recently introduced valence bond entropy~\cite{Alet07, Chhajlany07}. 
If one simply focuses on the IVB connecting the spin impurity with the rest of the system, 
we expect the probability $\cal{P}$ that the IVB has a length $r$ to decay like $\xi_K/r^2$
in the regime $r\gg \xi_K$, thus giving
\be
S_{\mathrm{ imp}}^{\mathrm{IVB}}(r)=\ln 2\left(1-\int_{1}^{r}{\cal{P}}{\rm{d}}r'\right)\sim A\ln(2)\frac{\xi_K}{r}~~(r\gg \xi_K).
\ee
Such a $1/r^2$ behavior is expected for a pure system~\cite{Alet09}.

Now consider the case of odd $R$.  The ground state now has a total spin of 1/2. 
We may again represent it by valence bonds but there are $(R-1)/2$ of them 
with one unpaired spin. This unpaired spin may or may not be the impurity spin. 
As $J_K'$ gets weaker it becomes more and more likely that the impurity 
spin is unpaired.  Clearly when $J_K'=0$ the other $R-1$ spins form a singlet 
leaving the impurity in a paramagnetic state. In this limit, for odd $R$, 
$S_{\mathrm{imp}}$ is precisely zero, due to the way we have defined it. Numerically, 
we find that the maximum in $S_{\mathrm{imp}}$, for odd $R$, occurs at a value of $J_K'$ 
such that $\xi_K\propto R$. This is due to trade-off between two competing effects. 
As we decrease $\xi_K$ from large values we increase the probability of having an 
IVB (i.e. of the impurity not being the unpaired spin).  This tends to 
increase $S_{\mathrm{imp}}$. However, once $\xi_K$ becomes less than $R$, the shortening 
of the IVB with decreasing $\xi_K$ becomes important and decreases $S_{\mathrm{imp}}$. 
It is now clear that the limits $\xi_K\to \infty$ and $R\to \infty$ don't commute. 
Taking $R\to \infty$ for any fixed $r$ and $\xi_K$ gives the same $S_{\mathrm{imp}}(r/\xi_K)$ 
as occurs for even $R$: a monotone decreasing function. However, holding $R$ fixed 
and varying $\xi_K$ gives a maximum of $S_{\mathrm{imp}}$ in the vicinity of $\xi_K\approx R$. 
Again, this is largely a heuristic picture. 

While we so far only have heuristic descriptions of $S_{\mathrm{imp}}$ when 
$\xi_K$ is of order $R$ and/or $r$, CFT methods can be used to calculate 
an analytic expression for $S_{\mathrm{imp}}$ in the opposite limit $\xi_K\ll r$ 
(for any ratio of $r/R$)~\cite{Sorensen07a,Sorensen07b}. (Note that in this limit the dependence on 
the parity of $R$ disappears.) This calculation is based on doing perturbation theory 
for $S_{\mathrm{imp}}$ in the leading irrelevant operator at the strong coupling, infrared fixed point. 
In fact such perturbation theory is very powerful and very well-known for the Kondo effect, 
going under the name of Nozi\`eres local Fermi liquid theory (FLT). It has been used long ago 
to calculate the leading dependence at low temperature of thermodynamic and transport 
quantities. This approach can be based on the continuum limit field theory of Eq. (\ref{Hspin}). 
The infrared stable, strong coupling fixed point Hamiltonian does not contain the impurity 
spin since it is screened and breaking this singlet costs an energy of $O(T_K)$. 
The low energy Hamiltonian at energy scales $\ll T_K$ only contains the continuum 
WZW fields. In this simple, spin-only model, the only effect of the Kondo 
interaction, once it has renormalized to strong coupling, is to switch the 
finite size spectrum of the WZW model between the $s=0$ and $s=1/2 $ conformal 
tower, corresponding to removing one site. To perturb around this fixed point 
we must identify the leading irrelevant operator which we expect to appear 
as a boundary interaction at $r=0$ only.  [It must be understood that 
this low energy theory is only valid at length scales large compared to $\xi_K$. 
We may think of the boundary interaction as being smeared over a distance of 
$O(\xi_K)$ but this is effective the same as being right at the boundary, $r=0$ 
in the effective Hamiltonian.] The leading irrelevant boundary operator, which must 
have $SU(2)$ symmetry, is simply $\vec J_L^2(r=0)$. [Recall that the BC 
$\vec J_L(0)=\vec J_R(0)$ means there is only one boundary current operator.]  
Very fortunately, this leading irrelevant interaction is proportional to 
the bulk energy density, as we see from Eq. (\ref{HsL}).  Defining this 
energy density:
\be {\cal H}_{s,L}(r)=
(v/6\pi ) \vec J_L\cdot \vec J_L \ee
the leading irrelevant interaction at the infrared stable fixed point is 
conventionally written:
\begin{equation}
H_{int}=-(\pi \xi _{K}){\cal H}_{s,L}(0).  \label{Hintcon}
\end{equation}
 This can be taken as a precise definition of the crossover length scale $\xi_K$. 
The fact that $H_{int}$ can be written in terms of ${\cal H}_{s,L}$ is 
very convenient for calculating the leading perturbation to the 
entanglement entropy because we may simply take over the earlier results 
of Calabrese and Cardy. These imply the leading correction:
\be S_{\mathrm{imp}}\to \pi \xi_K/(12r),\label{eq:FLT_FSS}\ee
representing first order perturbation theory in $\xi_K$, valid 
when $r\gg \xi_K$. This result is obtained for infinite $R$ but 
we may obtain the finite $R$ result by a standard conformal 
transformation:
\begin{equation}
S_{\mathrm{imp}}=\frac{\pi \xi _{K}}{12{R}}\left[ 1+\pi (1-\frac{r}{{R}})\cot (\frac{\pi r}{%
{R}})\right] .  \label{FLT_FSE}
\end{equation}
\begin{figure}[!ht]
\begin{center}
\includegraphics[width=10cm,clip]{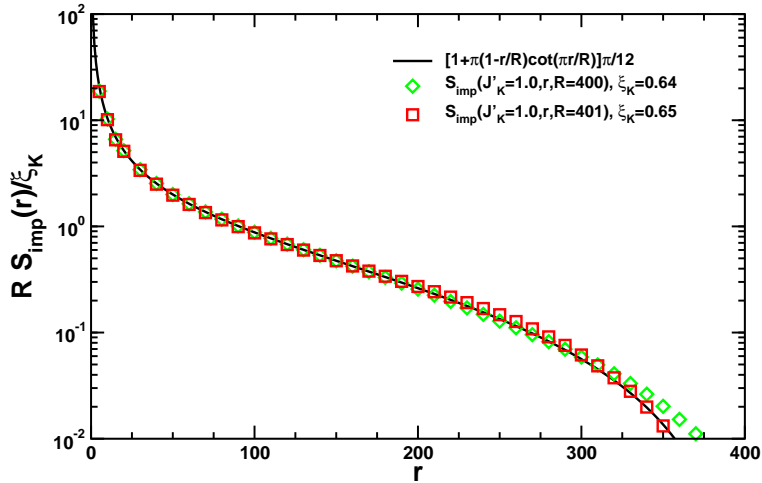}
\end{center}
\caption{DMRG results with $m=256$ for the $J_1-J_2$ chain. $S_{imp}(J^{\prime}_K=1,r,{R})$ for ${R}=400,401$ with $J_2=J_2^c$ compared to the
 FLT prediction, Eq.~(\ref{FLT_FSE}). }
\label{fig:simp1FLT}
\end{figure}
We find that this formula fits our numerical data, for both even and odd $R$, 
extremely well as illustrated in Fig.~\ref{fig:simp1FLT}.  Note that there are essentially no free parameters in the fit 
except for $\xi_K$ which can be determined independently as a function of $J_K'$ 
and is expected to behave exponentially at weak coupling:
\be \xi_K\propto \exp [1.38/J_K'].\ee
(The parameter, $1.38$ is {\it not} a fitting parameter but is rather determined 
from a careful mapping of the weak coupling of the end spin to the 
Kondo coupling in the usual fermion Kondo model.)
 This good fit of the CFT predictions to DMRG results on the 
spin chain version of the Kondo model is rather striking confirmation 
of the universality of entanglement entropy.

\subsection{Topological entanglement entropy}~\label{sec:topent}
\begin{figure}[!ht]
\begin{center}
\includegraphics[width=10cm,clip]{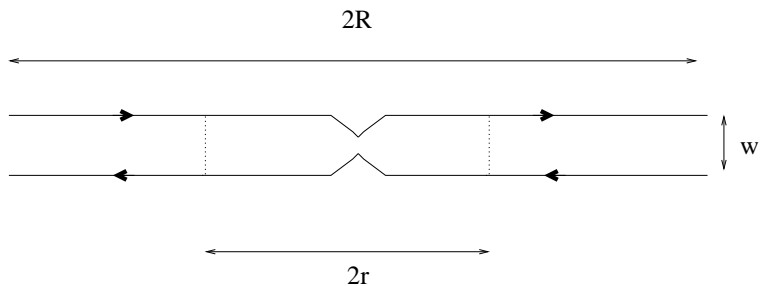}
\end{center}
\caption{A long thin Hall bar with a chiral edge current and a constriction. In sub-section 3.3 
we let $R\to \infty$ and 
consider the entanglement entropy for the region of width $2r$ with the rest of the 
Hall bar. In subsection 3.4 we take $R$ finite and consider the entanglement of the left side 
of the entire system with the right side.}
\label{fig:Hall}
\end{figure}

We now consider a class of gapped 2D systems known as topological insulators. 
These occur most famously as models for the fractional quantum Hall effect although 
other experimentally relevant possibilities have been conjectured. The gapped 
excitations of these systems can exhibit non-abelian statistics and are 
currently of great interest as possible topological quantum computers. 
It is rather difficult to demonstrate numerically that a given microscopic model 
has a topological ground state of a given type.  Recently, it was observed that 
this information can be extracted from the entanglement entropy. 
If we 
consider the entanglement entropy of a finite region  inside a  2D insulator of 
infinite extent then we expect ``area law'' behavior:
\be S=\alpha R'\ee
where $R'$ is the length of the perimeter of the region and $\alpha$ is a non-universal constant. 
However, it was recently shown \cite{Kitaev06,Levin06} that there is a sub-leading universal topological 
term in the entanglement entropy which is independent of the length and shape of the perimeter:
\be S=\alpha R'-\ln {\cal D}.\ee
To actually extract this term, it was proposed to divide the infinite 2D space into 
3 or 4 imaginary finite regions and 1 infinite one and to calculate a sum and difference of various 
entanglement entropies so that the term $\propto R$ cancels. The connection of this 
term with the impurity entropy, $\ln g$ was suggested in \cite{Kitaev06}.  (See also \cite{Levin06}). It was 
further elucidated in \cite{Fendley06} using the connection of a 2D topological 
insulator with a 1D edge model. Consider for example a ``quantum Hall bar'', a macroscopic sample 
of a topological insulator with edges. It is known that the electric current responsible for 
the Hall conductivity, $\sigma_{xy}$, flows around the edges only, in a clean sample, 
in a direction determined by the magnetic field direction, which is perpendicular to the plane. 
The low energy excitations on the edge correspond to a chiral CFT, meaning that the excitations 
are moving in only one direction. By considering a long thin hall bar, such as in Fig. (\ref{fig:Hall}) 
one can formally group together the right moving excitations on the upper edge and the left moving 
excitations on the lower edge to obtain a parity symmetric CFT. (As in all CFT's the left and 
right moving sectors are decoupled.) This is a particularly convenient formulation if a 
constriction is created in the Hall bar, corresponding to a narrowing of the bar (usually 
imposed with gate voltages) at one point, as shown in Fig. (\ref{fig:Hall}).  The constriction 
leads to ``back-scattering'' i.e. reflection of right-movers  approaching the constriction 
on the upper edge into left-movers leaving it on the lower edge. Integrating out the 
gapped bulk modes, this can be described by a purely 1D parity symmetric field theory: 
a CFT with a local back-scattering interaction.  In some situations, for example 
a simple CFT corresponding to spinless fermions with repulsive interactions, this 
back-scattering is a relevant interaction and can block all transport between left and 
right sides of the 1D system at low energies and long lengths. (This can be understood 
in terms of the boundary RG discussed in subsection 2.1). To make this connection more explicit 
let's consider the case where the constriction has zero width in the $x$-direction and is parity invariant 
so that it acts only on the parity even channel of excitations.  It is then 
possible to make a ``folding transformation'' mapping the 1D infinite system 
to a 1D semi-infinite system, $r\geq 0$ with the scatterer at $r=0$.  The incoming parity even excitations 
are mapped to left movers and the outgoing parity even excitations to right movers. The 
back-scattering interaction becomes a boundary sine-Gordon interaction for the boson field 
corresponding to the parity even excitations. When the back-scattering is relevant it 
leads to an RG flow from Neumann to Dirichlet BC's on the parity even boson. This is 
associated with a non-zero change in the impurity thermodynamic entropy, $\ln g$ which depends on 
the strength of the bulk excitations. 
In the original 2D model, we may think of the relevant constriction as effectively 
breaking the Hall bar into 2 pieces, each with its own chiral edge modes, as shown in Fig. (\ref{fig:Hall}). 

Again we may relate the impurity thermodynamic entropy to the impurity entanglement entropy. 
To do this, in the original 2D model, it is convenient to consider the entanglement of some 
region, $A$, containing the constriction and extending a distance $r$ on either side of it, 
with the rest of the infinite Hall bar. The corresponding entanglement entropy, in the 
1D folded formulation can be decomposed into a sum of contributions from even and odd 
parity modes.  Only the even parity part is affected by the constriction. The 
corresponding entanglement entropy for the even parity edge modes is:
\be S=\frac{1}{3}\ln \left(\frac{r}{a}\right)+\ln g+s_1\ee
the standard result for a finite region in an infinite CFT with $c=1$ (eg. with periodic BC's). 
Note that we have so far only considered the entanglement between region $A$ and the 
rest due to the gapless edge modes.  We expect additional contributions from the 
bulk modes, $2\alpha w$, where $w$ is the width of the Hall bar and hence $2w$ is the length of 
the boundary of region $A$.  Note that the extra term $(1/3)\ln r$ does not occur in 
the formulation of \cite{Kitaev06,Levin06} where the region $A$ does not include any physical 
boundaries and hence does not include any edge states. 

Now consider the change in entanglement entropy for region $A$ with the rest, as we increase 
the length, $2r$, of region $A$. We then expect an RG flow 
from the clean Hall bar at small $r$ to 
a Hall bar broken in two at large $r$, with the 
crossover length scale determined by the strength of the back-scattering. 
From the viewpoint of the low energy 1D model, the change results 
from the RG flow of the BC and is given by the change in $\ln g$:
\be S_{\mathrm{IR}}-S_{\mathrm{UV}}=\ln g_{\mathrm{IR}}-\ln g_{\mathrm{UV}}.\ee
[Note, in particular, that the $(1/3)\ln r$ term is the same in either 
perfectly transmitting or perfectly reflecting limits. In the first case 
it is the bulk 1D entanglement entropy for a region of length $2r$ in an infinite 
system.  In the second it is twice the entanglement entropy for a region of 
length $r$ in a semi-infinite system.]
  On the other hand, 
from the viewpoint of topological entropy, the constriction has broken the Hall bar 
into two pieces, without effecting the length of the perimeter, separating region $A$ from 
the rest, $2w$. This is expected to double the topological entropy:
\be S_{\mathrm{IR}}-S_{\mathrm{UV}}=-\ln {\cal D},\ee
since, in the infrared limit, we simply get twice the entanglement entropy of 
the Hall bar in the region $0<r'<r$ with the region $r'>r$,  $2(\alpha w-\ln {\cal D})$ 
with zero entanglement between left and right sides, through the constriction. 
Thus we conclude that:
\be \ln {\cal D}=\ln g_{\mathrm{UV}}-\ln g_{\mathrm{IR}}.\ee
Thus the topological entropy of a gapped 2D system is equal to the change in 
impurity entropy of the corresponding 1D gapless edge theory under a change 
in CIBC corresponding to breaking the system into two pieces. 
We note in passing that more generally, the local interactions associated with 
the constriction could lead to non-trivial boundary conditions not 
simply corresponding to perfect transmission or reflection. This would 
correspond to a different value of $g_{\mathrm{IR}}$ and probe other 
features of the topological insulator.

\section{Bulk impurity effects}\label{sec:peschel}
Another novel type of impurity entanglement entropy was studied in \cite{Levine04} 
and \cite{Peschel05,Zhao06}. We consider the same type of model as discussed in sub-section~\ref{sec:topent},
a Luttinger liquid with a back-scatterer, considering 
 the 1D formulation of the model. We now let the total system size $2R$ be finite 
with the back-scatterer in the centre. Rather than considering 
the entanglement of a central region containing the origin with the rest, 
these authors considered instead the entanglement entropy of the region to the 
left of the constriction with the region to the right.
  With no back-scattering this is:
\be S=\frac{c}{6}\ln \left(\frac{R}{a}\right).\ee
With relevant backscattering, the 1D wire is effectively broken into 
two disconnected parts at long length scales so we might expect the 
entanglement entropy to vanish asymptotically. This was studied 
numerically using DMRG in a critical Heisenberg XXZ spin chain with Hamiltonian:
\be H=\sum_jJ_j[(S^x_jS^x_{j+1}+S^y_jS^y_{j+1})+\Delta S^z_jS^z_{j+1}].\ee
For uniform couplings, this model is in a gapless Luttinger liquid phase for $-1<\Delta \leq 1$. 
Here $J_j=1$ for all links except for one in the middle of the chain 
where it has the value $J_{\mathrm{imp}}<$ with $0<J_{\mathrm{imp}}<1$. This weakened link is a relevant 
perturbation for $\Delta >0$ but irrelevant for $\Delta <0$. 
The DMRG results were consistent with $S/\ln R$ going to zero at large $R$ 
for $0<\Delta \leq 1$ but going to $(1/6)$, the expected $c=1$ value, 
for $-1<\Delta <0$. A quantitative theory of the $R$-dependence has not yet been developed, 
as far as we know.

Properties of the impurity entanglement in the $s=1/2$ XX chain in a transverse field where the field
strength is perturbed at a single central site have also been investigated~\cite{Apollaro06,Apollaro08}.

\section{The alternating part of the entanglement entropy}\label{sec:altent}
\subsection{Open boundary induced alternation}
As discussed in Ref.~\cite{Wang04,Laflorencie06}, in the case of AF spin chains with OBC, the von Neumann entropy can be written as a sum of two contributions:
\be
S(r,R)=S_{\rm U}(r,R)+(-1)^rS_{\rm A}(r,R).
\ee
\begin{figure}[!ht]
\begin{center}
\includegraphics[height=1.5cm,clip]{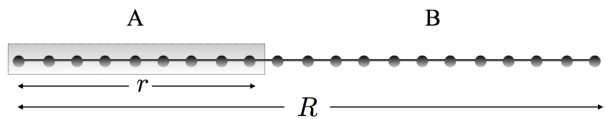}
\end{center}
\caption{Schematic picture of a chain of total length $R$ with OBC. The system has been cut in two parts A and B. Sub-system A, of length $r$, includes one open end.}
\label{fig:OBC_PICT}
\end{figure}
The uniform part $S_{\rm U}(r,R)$, in the case where the block of size $r$ contains one open end (depicted in Fig.~\ref{fig:OBC_PICT}), 
is given by the CFT result~\cite{Cardy04}, Eq.~(\ref{gL}):
\be
S_{\rm U}(r,R)=\frac{c}{6}\ln\left[\frac{2R}{a\pi}\sin\left(\frac{\pi r}{R}\right)\right] + \ln g +s_1/2,
\label{eq:OBCU}
\ee
\begin{figure}[th]
\begin{center}
\includegraphics[width=10cm]{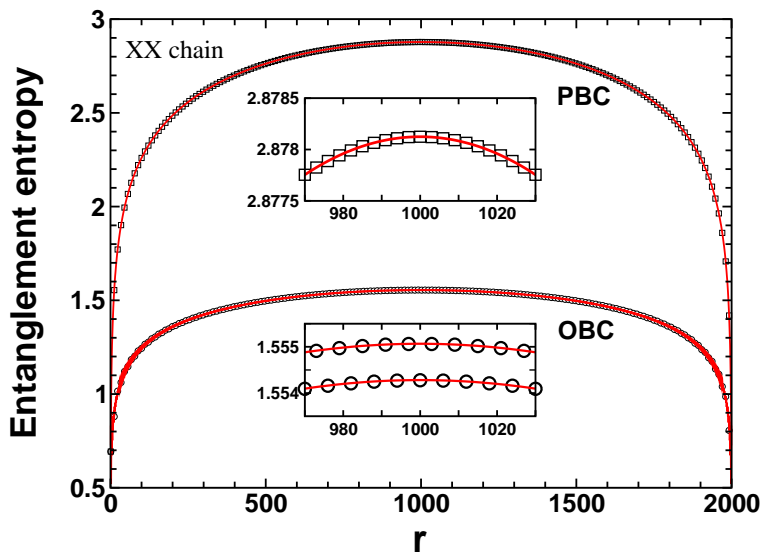}
\end{center}
\caption{Results from ED of XX chains with $R=2000$
  spins $1/2$. The von Neumann entropy $S(r,R=2000)$ is plotted vs the
  subsystem size $r$ for PBC (upper symbols) and OBC (lower
  symbols); the insets shows zooms around the chain center. The full
  lines are fits: Eq.~(\ref{eq:PBC}) with $s_1=0.726$ for PBC. In the OBC
  case, uniform and staggered terms $\frac{0.055-(-1)^r0.25}{\frac{R}{\pi }\sin
\left( \frac{\pi r}{R}\right)}$ have been added to Eq.~(\ref{eq:OBCU}). Reprinted from \cite{Laflorencie06}.}
\label{fig:XX}
\end{figure}
where $\ln g$ is the boundary entropy introduced in Ref.~\cite{Affleckg} and $s_1$ is a non-universal constant.  $S(r,R)$ can be exactly computed numerically using either DMRG or exact diagonalization (ED). For the $S=1/2$ XXZ chain,
\be
{\cal{H}}_{\rm xxz}=J\sum_{r=1}^{R-1}\left(S_{r}^{x}S_{r+1}^{x}+S_{r}^{y}S_{r+1}^{y}+\Delta S_{r}^{z}S_{r+1}^{z}\right),
\ee
which is critical for $|\Delta|\le 1$, there is a special point at $\Delta=0$
(XX point) where the spin chain is equivalent to free-fermions. There, one can
numerically compute the von Neumann entropy~\cite{Chung01} over for very large
systems using ED, as shown in Fig.~\ref{fig:XX} both for PBC and OBC with
$R=2000$ sites.  For periodic chains, the numerical results for the entropy are
very well described by the expression~\cite{Cardy04}, Eq.~(\ref{Spbc}):
\be S(r,R)=\frac{c}{3}\ln\left[\frac{R}{a \pi}\sin\left(\frac{\pi
    r}{R}\right)\right] + s_1, \label{eq:PBC} \ee
with $c=1$ and $s_1\simeq 0.726$ as predicted in Ref~\cite{Jin04} for
free-fermions.  On the other hand for OBC, besides the uniform logarithmic
increase Eq.~(\ref{eq:OBCU}), there are additional uniform and staggered terms
which decay away from the boundary $(0.055-(-1)^r 0.25)/\left[\frac{R}{\pi
}\sin \left(\frac{\pi r}{R}\right)\right]$. Not predicted by CFT, the origin of
the alternating term has been carefully investigated using ED and DMRG for
critical XXZ chains in \cite{Laflorencie06}. Such a phenomenon has also been
observed in several other cases where DMRG were applied for open systems. As
discussed above, in Ref.~\cite{Peschel05,Zhao06} Peschel and co-workers,
          studying the effect of interface defects in critical spin chains,
          reported the observation of such oscillations. Fermions or bosons
          confined in 1D geometries with OBC are also affected by such a
          modulation, as reported for
          fermionic~\cite{Legeza07,Szirmai08,Roux09} and
          bosonic~\cite{Lauchli08} Hubbard-like models. 

\begin{figure}[!ht]
\begin{center}
\includegraphics[width=8cm,clip]{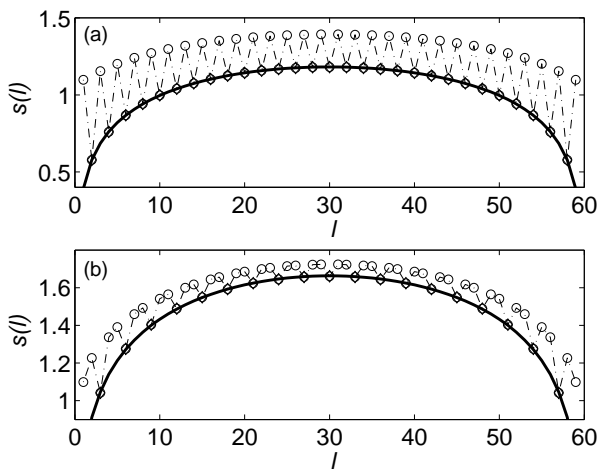}
\end{center}
\caption{von Neumann entropy computed by DMRG for the $S=1$ bilinear-biquadratic model Eq.~(\ref{eq:S1BB}) on open chains  for two critical points: (a) $\theta=-\pi/4$ and (b) $\theta=\pi/4$. Black lines are fits to the expression Eq.~(\ref{eq:OBCU}) with (a) $c=1.5$ and (b) $c=2$. Reprinted from \cite{Legeza07}.}
\label{fig:spin1}
\end{figure}
\begin{figure}[!ht] \begin{center}
\includegraphics[width=8cm,clip]{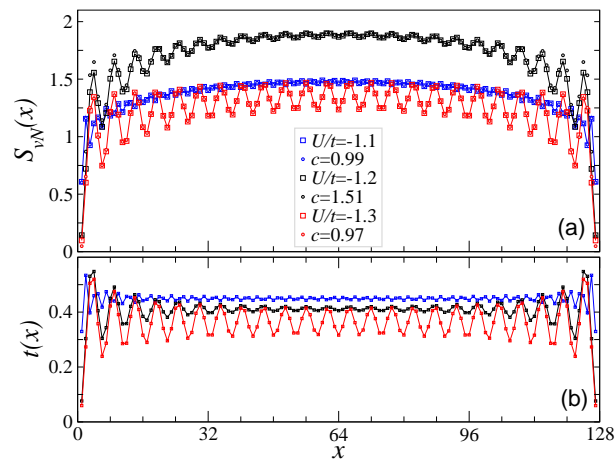} \end{center} \caption{Von
  Neumann block entropy $S(x)$ for a block of size $x$ (a) and local kinetic
    energy $t(x)$ (b) around the critical point of a  spin 3/2 fermionic
    Hubbard model with attractive interactions $U/t=-1.2$ for fixed $V/t=-2$
    and $n=0.75$ with $L=128$. Fits using the oscillations of the local kinetic
    energy term (b) are quite accurate, allowing for the determination of the
    central charge $c$. Reprinted from \cite{Roux09}.} \label{fig:Roux}
    \end{figure}

An interesting example of critical spin chain is the $S=1$ bilinear-biquadratic model
\be
{\cal{H}}=\sum_{r=1}^{R-1}\left[\cos\theta\left(\vec S_{r}\cdot\vec S_{r+1}\right)+\sin\theta\left(\vec S_{r}\cdot\vec S_{r+1}\right)^2\right],
\label{eq:S1BB}
\ee
which displays conformally invariant critical points at $\theta=\pi/4$ with
$c=2$ and at $\theta=-\pi/4$ with $c=3/2$. As studied by Legeza and
co-workers~\cite{Legeza07}, the dominant correlations at $k=\pi$ for
$\theta=\pi/4$ and $k=2\pi/3$ for $\theta=-\pi/4$ show up in the von Neumann
entropy as displayed in Fig.~\ref{fig:spin1}. However, a quantitative study of
this boundary-induced term was not achieved in Ref~\cite{Legeza07} where the
authors simply performed a fit to the expression Eq.~(\ref{eq:OBCU})
  restricting to the lower points: $r=2p$ ($\theta=-\pi/4$) and $r=3p$
  ($\theta=\pi/4$), with $p$ integer. Another interesting case has been studied
  with DMRG by Roux and collaborators \cite{Roux09} in the context of spin 3/2
  fermionic cold atom with attractive interactions confined in a 1D optical
  lattice. They also found OBC-induced non-uniform features in the von Neumann
  entropy $S(r,R)$ with $2 k_F$ oscillations that also appear in the local
  density and kinetic energy $t(r,R)$.  Oscillations of $S(r,R)$ and $t(r,R)$
  appear to be directly related as displayed in Fig.~\ref{fig:Roux}. Varying
  the parameters of the fermionic Hubbard model studied in this
  context~\cite{Roux09}, there is a critical point with $c=3/2$ which separates
  two phases with $c=1$. Imposing the non-uniform part of the von Neumann
  entropy to be directly proportional to the non-uniform part of the kinetic
  energy a good fit was obtained by Roux and collaborators with a clear jump in
  the central charge $c=1\to 3/2$ at the critical point~\cite{Roux09}.

%
\subsection{Valence bond physics}
Open boundary induced oscillations in the entanglement entropy appear to be a
quite general phenomenon, as also observed in a valence bond physics
framework~\cite{Alet07,Chhajlany07,Jacobsen08,Kallin09}. For SU(2) invariant
spin systems, the $S_{\rm tot}=0$ sector can be studied by quantum Monte Carlo
simulations in the valence bond basis~\cite{Sandvik05}. In such a framework,
            one can define a Valence Bond Entropy which displays surprising
            similarities with the von Neumann
            entropy~\cite{Alet07,Chhajlany07,Jacobsen08,Kallin09}.
\begin{figure}[!ht] \begin{center} \includegraphics[width=8cm,clip]{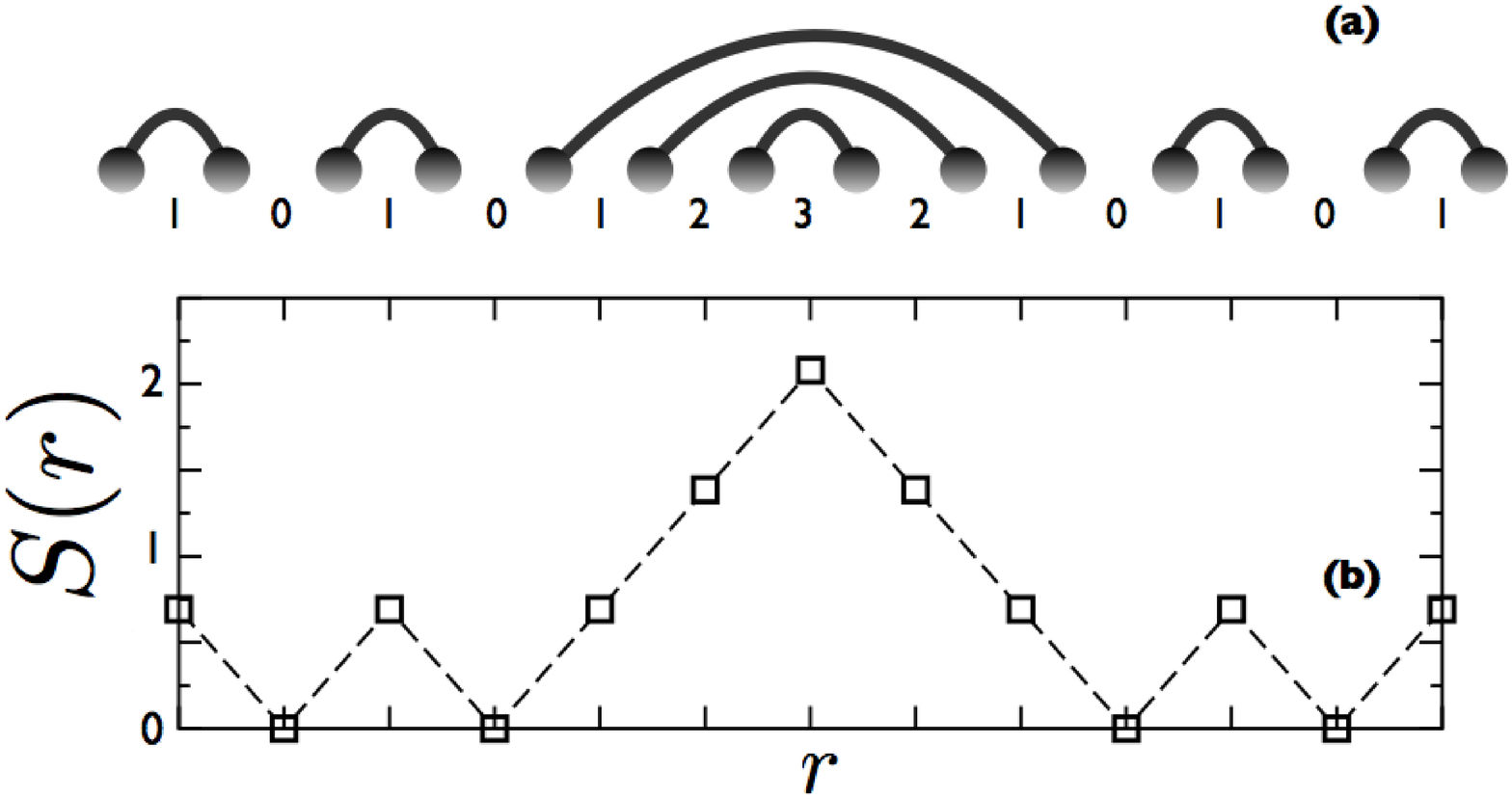}
\end{center} \caption{(a) Schematic picture for a particular valence bond state
  favored by OBC. The number of crossing bonds $N(r)$ is given. (b)
    Corresponding von Neuman entropy.} \label{fig:RVB} \end{figure}
For random bond spin chains, Refael and Moore~\cite{Refael04} achieved a very
nice calculation of the von Neumann entropy, simply observing that for a single
valence bond configuration (as depicted for instance in Fig.~\ref{fig:RVB}
    (a)), the entropy is just given by $N(r)\times \ln 2$, where $N(r)$ is the
number of singlet bonds crossing the interface between the two sub-systems.
Such an idea was successfully checked numerically in the random bond
case~\cite{Laflorencie05,Hoyos07,Alet07}. In the disorder free case, while the
ground-state is a highly non-trivial superposition of a huge number of valence
bond configurations, such a phenomenological approach appeared to be extremely
useful to understand the oscillating features~\cite{Laflorencie06}. Indeed,
       since the translational invariance is explicitly broken by the open
       ends, there is tendency towards dimerization near the open edges. Such
       an effect can be computed very precisely (see below) but already at a
       qualitative level, this short-range singlet formation in the vicinity of
       open boundaries can be interpreted as an alternation of \textit{strong}
       and \textit{weak} bonds along the chain, thus leading to a similar
       alternation of $S(r,R)$.  Indeed, the boundary spin at $r=1$ will have a
       strong tendency to form a singlet pair with its only partner on the
       right hand side. On the other hand the spin located at $r=2$ will be
       consequently less entangled with its right partner at $r=3$ since it
       already shares a strong entanglement with its left partner. A typical
       valence bon configuration favored by OBC is depicted in
       Fig.~\ref{fig:RVB} (a) with the corresponding entropy (b).  Such a
       qualitative interpretation in term of "weak-strong" modulation is
       directly related to OBC induced Friedel-like oscillations that one can
       investigate in a more quantitative way, as we do now.
\subsection{Entropy oscillation and dimerization for critical XXZ chains}
%
\begin{figure}[!ht] \begin{center} \includegraphics[width=14cm,clip]{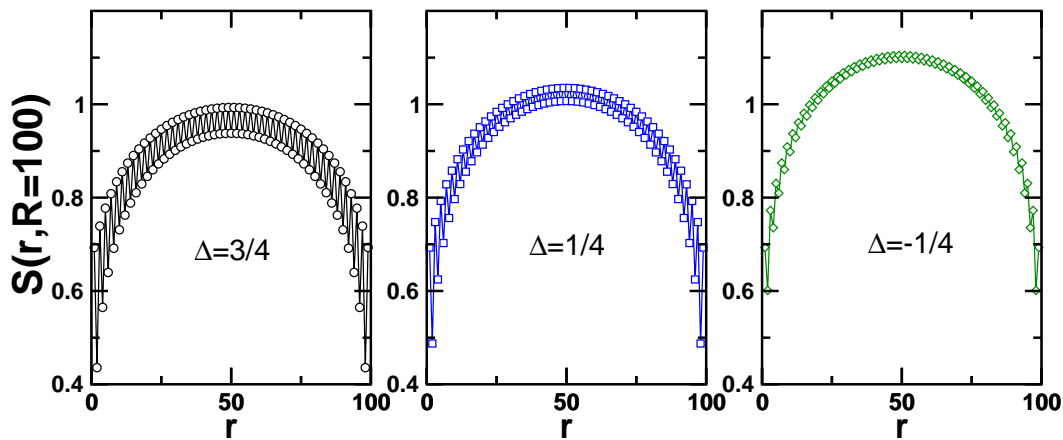}
\end{center} \caption{Von Neuman entropy of XXZ chains of lengths $R=100$ with
  OBC computed with DMRG for various anisotropies $\Delta$.} \label{fig:OBCXXZ}
  \end{figure}
\begin{figure}[ht!] \begin{center} \includegraphics[width=10cm]{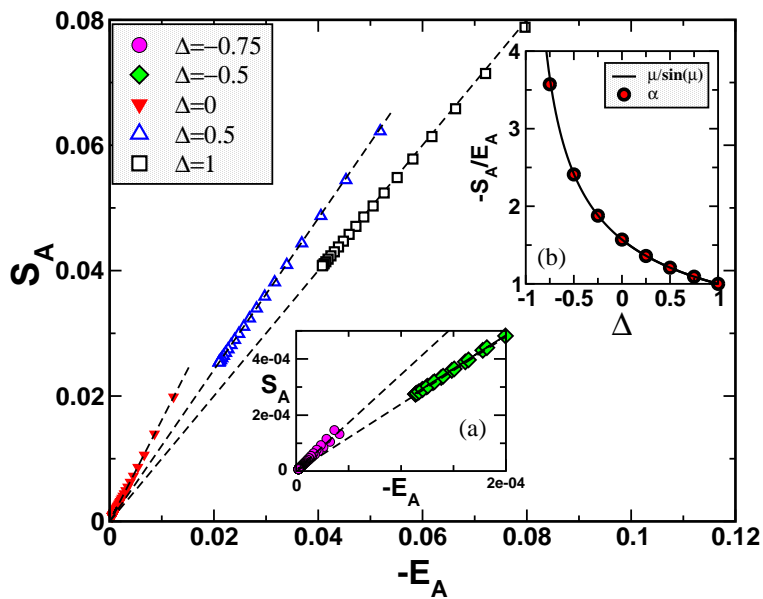}
\par \end{center} \caption{Linear behavior of the alternating part of the
  entanglement entropy, ${{S}}_A$ as a function of the alternating energy
    density, $-E_A$, both computed using DMRG  on critical open XXZ chains of
    size $200\le R\le 1000$ for a few values of the anisotropy $\Delta$. Data
    from ED at $\Delta=0$ are also shown for $R=2000$.  Dashed lines are linear
    fits of the form  ${{S}}_A=-\alpha E_A$ (see text).  The inset (a) is a
    zoom close to 0, showing data for $\Delta=-3/4$ and $-1/2$.  The inset (b)
    shows the prefactor $\alpha$ vs $\Delta$ extracted from the numerical data
    (circles), for a larger set of values of $\Delta$,  which is compared with
    $\pi /2v=\mu/\sin \mu$, $\mu =\cos^{-1}\Delta$. Reprinted
    from~\cite{Laflorencie06}.} \label{fig:Os-Dim} \end{figure}
\begin{figure}[ht!]
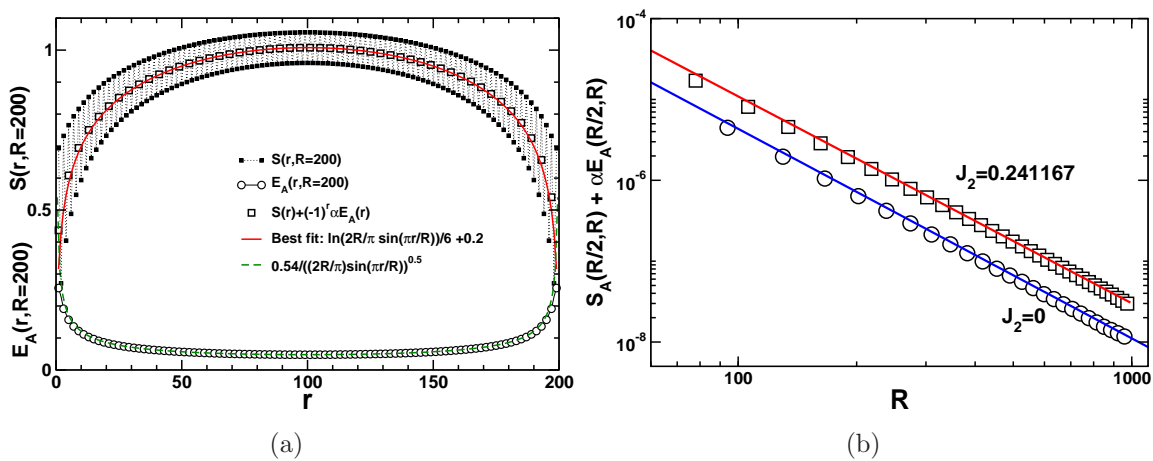
 \begin{center}
\subfigure[]{\includegraphics[width=0.48\textwidth,clip]{Dim}}
\subfigure[]{\includegraphics[width=0.48\textwidth,clip]{CompareSU2_Laflo06}}
\end{center} \caption{(a) Entanglement entropy $S(r)$ (black squares) and
  alternating part of the
    energy density $E_A(r)$ (open circles) computed with DMRG at $J_{2}^{c}$
    for
      $R=200$ sites. The dashed line is Eq.~(\ref{eq:EA}). The uniform part of
        $S(r)$, obtained by taking $S(r)+(-1)^r\alpha
        E_A(r)$ with $\alpha=1.001699$, is represented by open squares. The
        best fit,
             shown by a red curve, is indicated on the plot. (b) Comparison
               between the alternating part ${{S}}_A$  and $E_A$
               [Eq.~(\ref{eq:SE})] from DMRG with $m=512$ states, for
               Heisenberg models.  Power-law decay of ${S}_A(R/2,R)+\alpha
               E_A(R/2,R)$ drawn in a log-log plot, with $\alpha=1$ for the
               nearest neighbor chain ($J_2=0$) and $\alpha=1.00169$ at the
               critical second neighbor coupling $J_2=0.241167$.  Lines are
               power-law fits: $\sim R^{-2.56}$ for $J_2=0.241167$ and
               $R^{-2.59}$ for $J_2=0$. (a) and (b) are respectively reprinted
               from \cite{Sorensen07b} and \cite{Laflorencie06}.}
               \label{fig:CompareSU2} \end{figure}

The alternating part $S_A(r,R)$  has been studied in Ref.~\cite{Laflorencie06}
all along the critical regime of the XXZ chain $|\Delta|\le 1$. DMRG results
for $S(r,R)=S_{\rm U}(r,R)+(-1)^rS_{\rm A}(r,R)$ are shown in
Fig.~\ref{fig:OBCXXZ} for $R=100$ and various $\Delta$.  One sees immediately
that the oscillating part varies with $\Delta$ and decays faster in the
ferromagnetic regime. More quantitatively, it was shown~\cite{Laflorencie06}
that the alternating part $S_A(r,R)$ is directly proportional to the
alternating term in the energy density.  The energy density for XXZ spin
chains:
\begin{equation} \left\langle h_{r}\right\rangle =\langle
S_{r}^{x}S_{r+1}^{x}+S_{r}^{y}S_{r+1}^{y}+\Delta S_{r}^{z}S_{r+1}^{z}\rangle
\end{equation}
is uniform in periodic chains.  On the other hand, an open end breaks
translational invariance and there will be a slowly decaying alternating term
or ''dimerization'' in the energy density \begin{equation} \left\langle
h_{r}\right\rangle =E_{U}(r)+(-1)^{r}E_{A}(r), \end{equation} where $E_{A}(r)$
becomes nonzero near the boundary and decays slowly away from it. $E_{A}(r)$ is
obtained by abelian bosonization modified by OBC~\cite{Tsai00}. In the critical
region $\left| \Delta \right| \leq 1$, one
gets~\cite{Laflorencie06,Sorensen07b}
\be E_{A}(r,{\RR})\propto \frac{1}{[\frac{2{\RR}}{\pi }\sin (\frac{\pi
    r}{{\RR}})]^{K}}, \label{eq:EA} \ee
where $K$ is the Luttinger liquid parameter defined as $K=\pi /(2(\pi -\cos
      ^{-1}\Delta ))$ so that $K=1$ for an XX spin chain, $K=1/2$ for the AF
Heisenberg model, and $K\to\infty$ for the ferromagnetic Heisenberg case.
Based on DMRG data obtained~\cite{Laflorencie06} on critical open chains of
sizes $200\le R\le 1000$, we find a proportionality between $S_A$ and $E_A$.
More precisely, plotting ${{S}}_A$ as a function of $-E_A$ for various values
of the anisotropy $\Delta$ in Fig.~\ref{fig:Os-Dim}, we find a linear relation
${{S}}_A=-\alpha E_A$ with a prefactor perfectly described by $\alpha=\mu/\sin
\mu$, as shown in the inset of Fig.~\ref{fig:Os-Dim}, with $\mu =\cos
^{-1}\Delta$. We note that the velocity of excitations for the XXZ model is
given by $v=\pi (\sin \mu )/(2\mu )$ so that we may write this relation as 
\be {S}_A=-(\pi a^2/2v)E_A, \label{eq:SE} \ee
where we have introduced the lattice spacing, $a$, to make the entanglement
entropy a dimensionless quantity ($E_A$ has dimensions of energy per unit
    length.).
We emphasize that Eq. (\ref{eq:SE}) even holds for the Heisenberg model (with a
    proportionality coefficient $\alpha=1$) where both $E_A$ and ${S}_A$
display the same logarithmic corrections.  Indeed, at the Heisenberg point,
        $\Delta =1$, Eq. (\ref{eq:EA}) will have some logarithmic corrections
        due to the presence of a marginally irrelevant coupling constant in the
        low energy Hamiltonian, leading to~\cite{Laflorencie06,Sorensen07b}
        $E_A(r) \propto 1/[\sqrt{r}(\ln r)^{3/4}]$ in the limit $R\to \infty$.
        It is highly non-trivial to include both the log corrections and the
        finite size effects in $E_A(r,R)$. However, there {is} a simple result
        at $r=R/2$.  Including the cubic term in the $\beta$-function for the
        marginal coupling constant, and other higher order
        corrections~\cite{Barzykin99}, this becomes:
\be E_A(\frac{R}{2},R)=a_0  \frac{1+a_{2}/[\ln (R/a_{1})]^{2} }{\sqrt{R}[\ln
  (R/a_{1})+(1/2)\ln \ln (R/a_{1})]^{\frac{%
    3}{4}}}  ,  \label{logcorr} \ee
where $a_0$, $a_1$ and $a_2$ are constants. If one allows a frustrating second
neighbor coupling $J_2$ in the chain, at $J_2^c=0.241167 J$, this model is at
the critical point between gapless and gapped spontaneously dimerized phase and
the marginal coupling constant, and hence the log corrections are expected to
vanish here. In both cases ($J_2=0$ and $J_2=J_2^c$), we found proportionality
between $S_A$ and $E_A$, as shown in Fig.~\ref{fig:CompareSU2}.  The sum
$E_A(R/2,R)+\alpha{S}_A(R/2,R)$ is found to rapidly decay as a power-law, with
a power $\sim 2.5$ (see Fig.~\ref{fig:CompareSU2}).  Interestingly, for
$J_2=0.241167$, again linearity is observed, but with a prefactor $\alpha\simeq
1.001689$ not related to the spin velocity, $v$, which we have determined to be
$v\simeq 1.174(1)$~\cite{Laflorencie08}.  Hence,  Eq.~(\ref{eq:SE}) does not
hold in this case.
%

We emphasize that this alternating term 
in ${S}(r,R)$ is universal and 
should {\it not} be regarded as a correction 
due to irrelevant operators.   First of all, it is not a ``correction'', 
since it is alternating. Secondly, it decays with the same power law
as $E_A(r,R)$ which is seen to be a property of the fixed point, 
not the irrelevant operators. (However, for the Heisenberg model, 
$\Delta =1$, the log factor in $E_A(r,R)$ {\it is} due to the 
marginally irrelevant operator.)  The presence of a universal  
alternating term in ${S}(r,R)$ 
is connected with the antiferromagnetic nature 
of the Hamiltonian (not appearing, for example, in the quantum Ising 
chain~\cite{Zhou06})
and {\it does not} seem to follow from the general CFT 
treatment in \cite{Cardy04}. An analytic derivation of this 
phenomena remains an open problem. 

\subsection{Spin chain Kondo model}
\begin{figure}[!ht]
\begin{center}
\includegraphics[width=\columnwidth,clip]{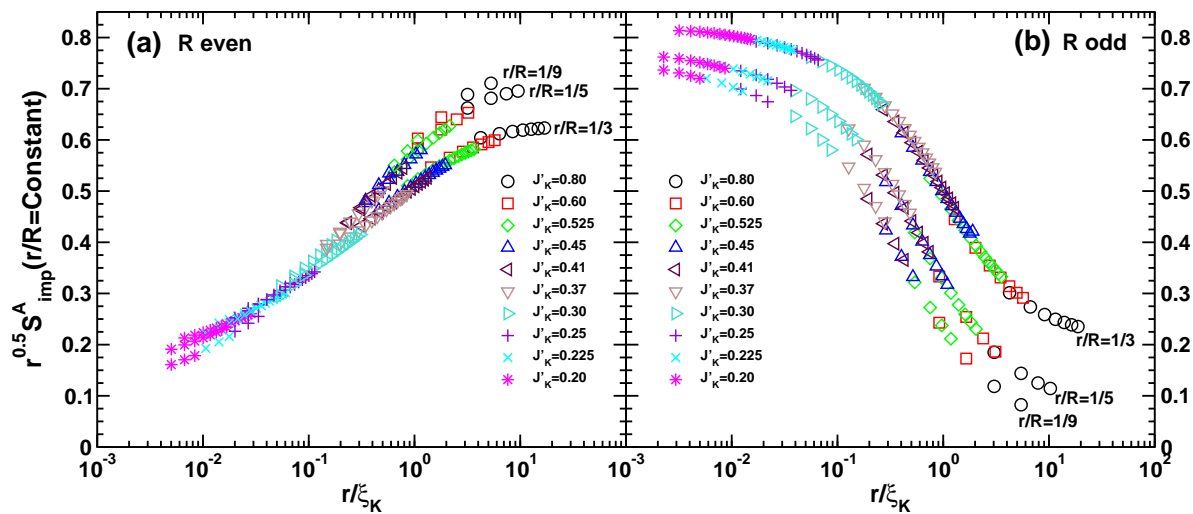}
\end{center}
\caption{$\sqrt{r}S^A_{imp}$ for fixed $r/R$ and a range of couplings $J'_K$ at $J_2^c$. (a) for $R$ even and (b) for $R$ odd.
The values of $\xi_K$ used are
obtained from the scaling of $S_{\mathrm{imp}}$ achieved in Fig.~\ref{fig:kondoDMRG}.
DMRG results with $m=256$ states. Reprinted from Ref.~\cite{Sorensen07b}.}
\label{fig:ASimprovr}
\end{figure}

We now turn to the spin chain Kondo model Eq. (\ref{eq:spinch}). In previous sections, only the uniform part of the impurity part of the von Neumann entropy has been investigated. Here we focus on the alternating part which is also present for impurity problems. We want to isolate the impurity contribution in the alternating part of $S(r,R)$. Following our fundamental definition of
\begin{figure}[!ht]
\bc
\includegraphics[width=\columnwidth,clip]{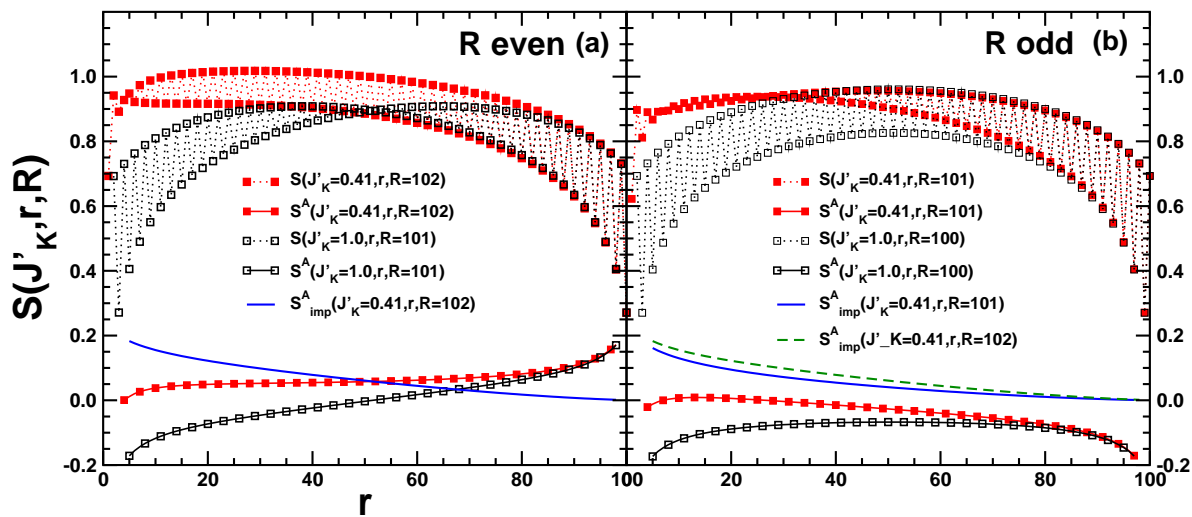}
\ec
\caption{(a) DMRG results for the total entanglement entropy, $S(J'_K,r,R)$ for a 102 site spin chain at $J^c_2$,
  with a $J'_K=0.41$ Kondo impurity ($\blacksquare$) along with $S(J'_K=1,r-1,R-1)$ ($\square$).
    For both cases is the extracted alternating part shown along with the resulting $S^A_{imp}(J'_K=0.41,r,R=102)$ for $R$ even.
(b) DMRG results for the total entanglement entropy, $S(J'_K,r,R)$ for a 101 site spin chain at $J^c_2$,
  with a $J'_K=0.41$ Kondo impurity ($\blacksquare$) along with $S(J'_K=1,r-1,R-1)$ ($\square$).
    For both cases is the extracted alternating part shown along with the resulting $S^A_{imp}(J'_K=0.41,r,R=101)$ for $R$ odd.
    For comparison we also show $S^A_{imp}(J'_K=0.41,r,R=102)$ for $R$ even from panel (a) (dashed line). Reprinted from Ref.~\cite{Sorensen07b}.}
\label{fig:RawASimp}
\end{figure}
$S_{\mathrm{imp}}$, Eq.~(\ref{Simpdef}), it is also possible to define the alternating part of the impurity
entanglement entropy:
\begin{equation}
S^A_{\mathrm{imp}}(J'_K,r,{\RR})\equiv S_A(J'_K,r,{\RR})-S_{A}(1,r-1,{\RR}-1).\label{ASimpdef}
  \end{equation}
As before, we have subtracted $S_A$ when the impurity is absent,
in which case both $r$ and $\RR$ are reduced by one and
the coupling at the end of this reduced chain, linking site $2$ to
$3$ and $4$, has unit strength. Applying this definition to numerical data involves some subtleties.
First of all $S^A$ is only defined up to an overall sign. Secondly, when calculating $S_{A}(1,r-1,{\RR}-1)$
we define this as $-S_{A}(1,r',{\RR}')$ with $R'=R-1$ since the shift from $r$ to $r-1$ implies a sign change in the
alternating part. For convenience we have therefore always exploited this degree of freedom to use a sign convention
that makes the resulting $S^A_{\mathrm{imp}}$ positive in all cases. In Fig.~\ref{fig:RawASimp} we show data for the
total entanglement entropy along with the extracted alternating parts and the resulting $S^A_{imp}$ for both
$R=102$ even and $R=101$ odd.  As was the case for the uniform part of $S_{\mathrm{imp}}$ (Fig.~\ref{fig:kondoDMRG}) we do not observe any special features in $S^A_{\mathrm{imp}}(r)$ for fixed $R, J_K'$ associated with the length
scale $\xi_K$ and in all cases $S^A_{imp}$ decays monotonically with $r$.
On the other hand, a possible scaling form for $S^A_{\mathrm{imp}}$ was suggested in Ref.~\cite{Sorensen07b}.
For $J_K'=1$ we have seen above that the alternating part of
the entanglement entropy, $S_A$, is proportional to the alternating part in the energy, $E_A$ which, for $J_2=J_2^c$ can be written as follows: $E_A(r)=f(r/R)/\sqrt{r}$ for some scaling function $f$. 
A generalization of the above formula to the case $J_K'\neq1$
imply that $S^A_{\mathrm{imp}}\sqrt{r}$ should be a scaling function, $f(r/R,r/\xi_K)$.
DMRG results for $S^A_{\mathrm{imp}}\sqrt{r}$ for fixed $r/\RR$ are shown in Fig.~\ref{fig:ASimprovr} for a range
of $J'_K$ and $\RR$. The values for $\xi_K$ used to attempt the scaling are the ones previously used for the scaling of the uniform part (Fig.~\ref{fig:kondoDMRG}).
Clearly the results for $S^A_{\mathrm{imp}}\sqrt{r}$ follow the expected scaling form. 

\section{Impurity entanglement in gapped spin chains}\label{sec:gap}
In previous sections the focus has largely been on work considering critical systems. However,
impurity entanglement in systems with a gap, corresponding to massive field theories with a
finite correlation length $\xi$, have also
been considered.  In this case the generalization of Eq.~(\ref{g}) to one dimensional systems with
a gap becomes~\cite{Cardy04}:
\begin{equation}
S(r)\sim \frac{c}{6}\ln \left(\frac{\xi}{a}\right).\label{eq:Sgap}
\end{equation}
So far, mainly two models have been studied, $s=1$ chains~\cite{Fan04,Fan06}  at the AKLT point~\cite{AKLT87,AKLT88} and $s=1/2$
$J_1-J_2$ chains~\cite{Sorensen07b} at the Majumdar-Ghosh point, $J_2=J_1/2$~\cite{MG70}. While the $s=1$ spin chain at the AKLT
point has a correlation length $\xi=1/\ln(3)$, the spin correlations in the $s=1/2$ chain at the MG point do not extend beyond
nearest neighbor and the correlation length is effectively zero.

\subsection{The $s=1$ ALKT chain}
Boundary effects in the entanglement entropy of a $s=1$ chain was studied in~\cite{Fan06} building on earlier work~\cite{Fan04}.
The model considered was the antiferromagnetic $s=1$ Heisenberg chain including a bi-quadratic term:
\begin{equation}
H=\sum_{j=-N_l+1}^{L+N_r-1}\left[{\vec S}_j\cdot{\vec S}_{j+1}+\frac{1}{3}({\vec S}_j\cdot{\vec S}_{j+1})^2\right].\ \ s=1.
\end{equation}
For this special value of the bi-quadratic coupling, termed the AKLT point, the ground-state is known exactly~\cite{AKLT87,AKLT88}
and this fact was exploited to perform exact calculations of the entanglement entropy. The entanglement of a sub systems consisting of
the {\it central} section of the chain,
$1\ldots L$ with the remaining {\it left and right} parts of the chain was considered. Here $N_l$ and $N_r$ describe the size
of the left and right parts of the system and it was shown~\cite{Fan06} that the boundary effects in the entanglement entropy
decay exponentially fast with $N_l,N_r$ on a length scale equal to the correlation length $\xi=1/\ln(3)$. Hence, impurity effects
in the entanglement impurity are in this case rather minor. In contrast, by considering a different subsystem impurity entanglement
in the Majumdar Ghosh model can be rather pronounced.

The partial concurrence of the two effective $s=1/2$ spins at the end of an open $s=1$ chain has also been studied~\cite{Venuti06}.

\subsection{The Majumdar Ghosh Model}
\begin{figure}[!ht]
\begin{center}
\includegraphics[height=3.5cm,clip]{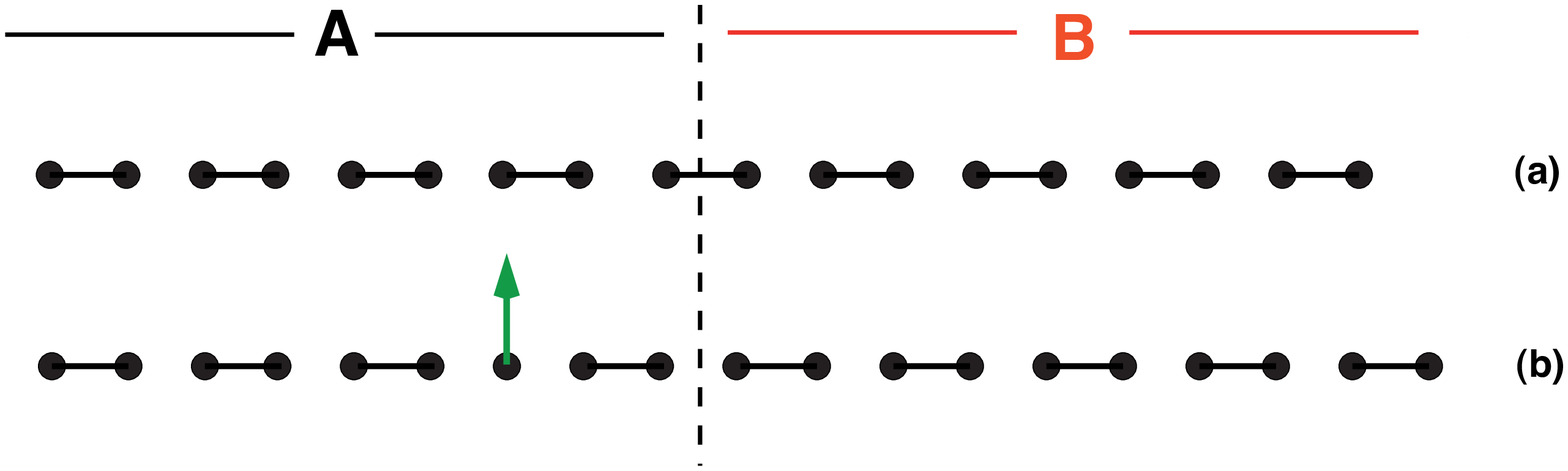}
\end{center}
\caption {MG state divided in two regions $A$ and $B$. (a) The total number of site $R$ is even and the ground-state is a singlet. (b)
${\color{green}\uparrow}$.  Reprinted from~\cite{Sorensen07b}}
\label{fig:MG}
\end{figure}
At the special point $J_2=J_1/2$, often referred to as the Majumdar-Ghosh~\cite{MG69a,MG69b,MG70} (MG) point,
the spin chain model Eq.~(\ref{eq:spinch}) with $J_K'=1$ is exactly solvable. The spin chain
has a gap and for $R$ {\it even}, in the presence of periodic boundary conditions, a two-fold degenerate gound-state of nearest neighbor dimers
either between sites $2n+1$ and $2n+2$ or $2n+2$ and $2n+3$, $n\ge 0$, with energy $E=-3RJ/8$. 
With open boundary conditions, for $R$ {\it even}, the ground-state is non-degenerate with dimers between sites $2n+1$ and $2n+2$, $n=0\ldots (R-1)/2$, and with
the same energy as the periodic case, $E=-3RJ/8$. See Fig.~\ref{fig:MG}(a).
If one now instead considers $R$ {\it odd}, an exact form for the ground-state wave-function and energy is not known but a
very precise variational form can be developed~\cite{SS81,Caspers82,Caspers84,Sorensen98,Uhrig99,Sorensen07b,Doretto09}.
For $R$ odd it is natural to 
consider states of the following form:
\begin{equation}
|n\rangle\equiv|\overbrace{-\ \ldots\ -}^n\ \uparrow\ -\ \ldots\ -\rangle.
\end{equation}
Here, $-$ indicates a singlet between site $r$ and $r+1$ and
$n=0,\ldots N_d-1=(R-1)/2$ refers to the number of dimers to the left of the soliton,
with $N_d$ the total number of dimers. See Fig.~\ref{fig:MG}(b).
Such states are often called {\it thin soliton} states (TS-states) because the soliton
resides on a single site and is not `spread' out over several sites as would have been the case
if one had included states with valence bonds longer than between nearest neighbor sites.
Note that the soliton only resides on the odd sites of the lattice, $r=1,3,\ldots , R$.
It is important to realize that the TS-state as defined are not orthornormal and even though
it is straight forward to form linear combinations $|m\rangle$ that are orthogonal~\cite{Uhrig99,Doretto09}
through the transformation 
$|m=0\rangle\equiv |n=0\rangle,\ \ |m\rangle\equiv (2/\sqrt{3})\left[|n\rangle+(1/2)|n-1\rangle\right]\ m\ge 1$,
one finds that for the purpose of calculating the entanglement this orthogonalization is less useful at the initial
stage of calculating the entanglement entropy.

If only the $^\Uparrow$ state of the gound-state doublet for $R$ odd is considered
one can write a thin soliton ansatz (TS-ansatz) for the ground-state wavefunction:
\begin{equation}
|\Psi_{TS}^\Uparrow\rangle\simeq \sum_{n=0}^{N_d}\psi^{sol}_n|n\rangle.
  \label{eq:TSansatz}
\end{equation}
It can be shown that the restriction to the $^\Uparrow$ state is not important since
any linear combination of the degenerate $^\Uparrow$ and $^\Downarrow$ states will 
yield the same entanglement entropy~\cite{Sorensen07b}.
The components of the ground-state wavefunction, $\psi^{sol}_n$ can be obtained in
a variational manner or through a simple analytical estimate~\cite{Sorensen07b}:
\begin{equation}
\psi^{sol}_n\simeq \sqrt{\frac{2}{N_d+2}}\sin\left(\frac{\pi(2n+2)}{\RR+3}\right).  \label{eq:sinform}
\end{equation}
With the $\psi^{sol}_n$ determined we can proceed with a calculation of the entanglement entropy. 
Due to the dimerization of the ground-state one sees that for $R$ even (and open boundary conditions) 
$S(r,R)$ simply oscilates with $r$ between $\ln(2)$ and 0 depending on whether $r$ is a site at the
beginning or end of a dimer, respectively. With a $J_K'\neq 1$ in Eq.~(\ref{eq:spinch}) the entanglement
is much richer and exact results are not available, however, {\it very precise} variational calculations~\cite{Sorensen07b} of $S(r,R)$
are possible. Let us first consider $R$ odd and $J_K'=1$ with $r$ odd. We first divide the system in
two parts ($A,B$) at $r$ and let $|\phi\rangle$ denote a basis for $1\ldots r$ and $|\psi\rangle$ a basis for $r+1..R$.
We then have:
\begin{equation}
|\Psi_{TS}^\Uparrow\rangle=\sum_{i,j=0}^{3}C_{i,j}|\psi_i\rangle|\phi_j
\rangle.\label{eq:gsodd}
\end{equation}
The above form follows since the system was divided in two parts at $r$ and since $r$ {\it is odd}. If the soliton
is to the left of $r$ and since $r$ is odd the division between $A$ and $B$ will not 'cut' a dimer. However, if
the soliton is to the right of $r$ the division will cut a dimer and we effectively get an additional soliton at
the end of the $A$ space leading to the following 3 separate cases
for $|\psi_i\rangle$:
\begin{eqnarray}
|\psi_1\rangle&=&\sum_{n=0}^{\frac{r-1}{2}}\psi^{sol}_n|\overbrace{-\
  \ldots\ -}^n\ \uparrow\ -\ \ldots\ -\rangle  \nonumber \\
|\psi_2\rangle&=&|\overbrace{-\ -\ -\ \ldots\ -\ -\ -\ }^{\frac{r-1}{2}}\
\uparrow\rangle  \nonumber \\
|\psi_3\rangle&=&|\overbrace{-\ -\ -\ \ldots\ -\ -\ -\ }^{\frac{r-1}{2}}\
\downarrow\rangle.
\end{eqnarray}
With the corresponding definitions for $|\phi_j\rangle$:
\begin{eqnarray}
|\phi_1\rangle&=&|\overbrace{-\ -\ -\ \ldots\ -\ -\ -\ }^{\frac{\RR-r}{2}}\rangle  \nonumber \\
|\phi_2\rangle&=&\sum_{n=0}^{\frac{\RR-r}{2}-1}
\psi^{sol}_{\frac{r+1 }{2}+n}|\downarrow\ \overbrace{-\ \ldots\ -}^n\ \uparrow\ -\ \ldots\ -\rangle \nonumber \\
    |\phi_3\rangle&=&\sum_{n=0}^{\frac{\RR-r}{2}-1}\psi^{sol}_{\frac{r+1}{2}+n}|\uparrow\ 
    \overbrace{-\ \ldots\ -}^n\ \uparrow\ -\ \ldots\ -\rangle.  
\label{eq:phij}
\end{eqnarray}
With these definitions it immediately follows that:
\begin{equation}
|\Psi_{TS}^\Uparrow\rangle= |\psi_1\rangle|\phi_1\rangle+\frac{1}{\sqrt{2}}%
\left[|\psi_2\rangle|\phi_2\rangle-|\psi_3\rangle|\phi_3\rangle\right].\label{eq:gsodd2}
\end{equation}
from which the coefficients $C$ defined in Eq.~(\ref{eq:gsodd}) can be determined.
The states $|\phi_i\rangle$ and $|\psi_j\rangle$ are clearly not orthonormal, however, by explicitly orthonormalizing
the states $|\psi_j\rangle$ region $B$ can be traced out and the reduced density matrix $\rho$ for region $A$ determined
It is straight forward to generalize this approach to the case where $r$ is even (and $R$ is odd with $J_K'=1$). 
\begin{figure}
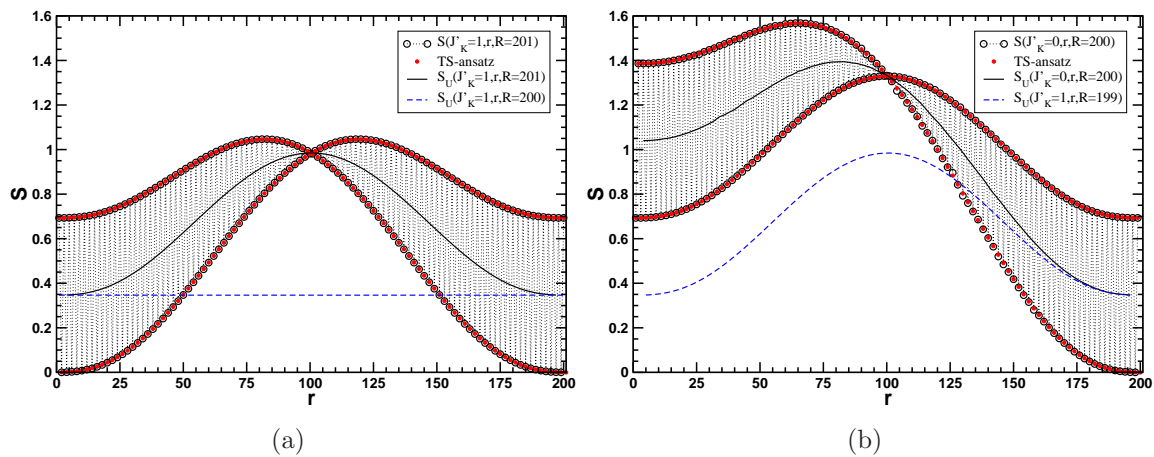

\subfigure[]{\includegraphics[width=0.48\textwidth,clip]{SSansatz}}
\subfigure[]{\includegraphics[width=0.48\textwidth,clip]{SSansatzJK0}}
\caption{(a)DMRG results with$m=256$ for $S(J^{\prime}_K=1,r,\RR=201)$ for the
spin-chain model with $J_2=1/2$ (MG model). The small solid circles represent the
theoretical result obtained using the TS-ansatz. Also shown are the uniform parts $S_U(J^{\prime}_K=1,r,\RR=201)$ and
$S_U(J^{\prime}_K=1,r,\RR=200)$ the difference of which is $S_\mathrm{imp}(J'_K=1)$.
(b)DMRG results with$m=256$ for $S(J^{\prime}_K=0,r,\RR=200)$ for the
spin-chain model with $J_2=1/2$ (MG model). The small solid circles represent the
theoretical result obtained using the TS-ansatz. Also shown are the uniform parts $S_U(J^{\prime}_K=0,r,\RR=200)$ and
$S_U(J^{\prime}_K=1,r,\RR=199)$ the difference of which is $S_\mathrm{imp}(J'_K=0)$. }
\label{fig:SSansatz}
\end{figure}
Results from such a calculation is shown in Fig.~\ref{fig:SSansatz}(a) where they are compared to DMRG results. 
Excellent agreement is observed. 
For comparison, results for $S_u(J^{\prime}_K=1,r,\RR=201)$ and $S_U(J^{\prime}_K=1,r,\RR=200)$ are also shown in Fig.~\ref{fig:SSansatz}(a).
The difference of these two uniform parts yields $S_{\mathrm{imp}}$.
The influence of the impurity spin coupled with $J_K'=1$ is clearly visible and extends over the {\it entire} range of $r$.

As already outlined in section~\ref{sec:boundent}, the entanglement of the impurity spin with the bulk of the
chain is sizable (one might even say maximal) even when $J_K'=0$. With some additional algebra it is possible
to extend the thin-soliton approach also to this case~\cite{Sorensen07b} by considering a decoupled impurity entangled with
a bulk chain of {\it odd} length $R-1$. Hence, the TS-ansatz can be applied to the bulk chain and precise variational
results obtained. It is crucial to note that the impurity spin and the bulk chain form a singlet state even though
their coupling, $J_K'$, is zero. That is, we're considering the $J_K'\to 0$ limit of the entanglement entropy. 
Results obtained from the TS-ansatz are shown in Fig.~\ref{fig:SSansatz}(b) where they are compared to DMRG results obtained using spin-inversion.
The DMRG calculations are performed on systems where the impurity spin is explicitly included resulting in a
4-fold degenerate ground-state consisting of a singlet and a triplet. The application of
spin-inversion symmetry is therefore necessary in order to distinguish the singlet state of interest from the degenerate triplet
state. The agreement between the numerical and variational results in Fig.~\ref{fig:SSansatz}(b) is excellent, the small discrepancies
visible at a few values of $r$ are due to complications using spin-inversion~\cite{Sorensen98b} in the DMRG calculations specific to this value of $J_2$.
For comparison, results for $S_u(J^{\prime}_K=0,r,\RR=200)$ and $S_U(J^{\prime}_K=1,r,\RR=199)$ are also shown in Fig.~\ref{fig:SSansatz}(b).
The difference of these two uniform parts yields $S_{\mathrm{imp}}$ which is non-zero throughout the system.

\begin{figure}
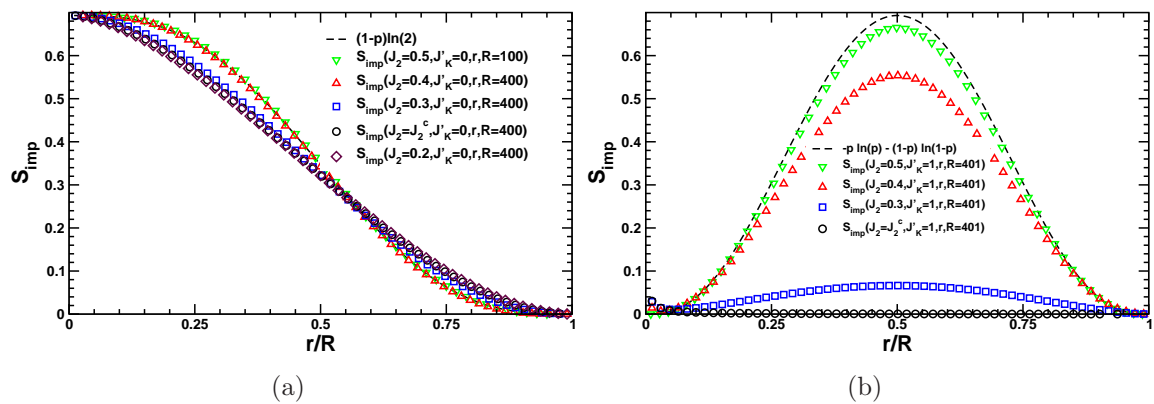

\subfigure[]{\includegraphics[width=0.48\textwidth,clip]{Simp0Scal}}
\subfigure[]{\includegraphics[width=0.48\textwidth,clip]{Simp1Scal}}
\caption{(a)DMRG results with$m=256$ for $S(J^{\prime}_K=0,r,\RR=400)$ for the
spin-chain model with $J_2/J=0.2,J_2^c,0.3,0.4,0.5$ shown along with the result for
$S_{\mathrm{imp}}^{\mathrm{IVB}}=(1-p)\ln(2)$ with $p$ calculated at $J_2/J=0.5$ (dashed line).
(b)DMRG results with$m=256$ for $S(J^{\prime}_K=1,r,\RR=401)$ for the
spin-chain model with $J_2/J=J_2^c,0.3,0.4,0.5$ . The dashed line represents 
$S_{\mathrm{imp}}^{\mathrm{SPE}}=-p\ln(p)-(1-p)\ln(1-p)$ with $p$ calculated at $J_2=J/2$.
Reprinted from~\cite{Sorensen07b}.
}
\label{fig:SimpScal}
\end{figure}

Perhaps surprisingly, it is possible to obtain a much more intuitive picture and yet still very precise by simply
assuming that the states $|\phi_i\rangle$ and $|\psi_j\rangle$ are orthogonal. With this assumption it is easy
to see from Eq.~(\ref{eq:gsodd2}) that the reduced density matrix for region $A$ (with $R$ odd) is simply:
\begin{equation}
    \rho= \left( \begin{array}{ccc} p&0 & 0\cr 0&
        \frac{1-p}{2} & 0\cr 0& 0& \frac{1-p}{2} \cr \end{array}\right),\ r\
      \mathrm{odd},
        \ \ \ \ \
\rho= \left( \begin{array}{ccc} 1-p&0 & 0\cr 0&
\frac{p}{2} & 0\cr 0& 0& \frac{p}{2} \cr \end{array}\right),\ r\ \mathrm{ even}.
\end{equation}
Here $p=\sum_{n=0}^{(r-1)/2}|\psi^{sol}_n|^2\sim r/R-\sin(2\pi r/R)/(2\pi)$ is the probability of finding the soliton in region $A$. From these
expressions for $\rho$ the entanglement entropy can easily be evaluated. In order to calculate $S_{\mathrm{imp}}$
the uniform part of $S$ for an even length chain with $J_K'=1$ is needed, but, as mentioned previously, this is simply $\ln(2)/2$ and
it follows that:
\begin{equation}
S_{\mathrm{imp}}(J^{\prime}_K=1,r,\RR)=-p\ln( p) -(1-p)\ln(1-p)\equiv S_{\mathrm{imp}}^{\mathrm{SPE}}, \RR\
    \mathrm{odd}.  \label{eq:sfpodd}
    \end{equation}
In this case the impurity entanglement arises solely from the entanglement of a single particle (the soliton) that
is present in the ground-state and one therefore refers to this contribution as the single particle entanglement, $S_{\mathrm{imp}}^{\mathrm{SPE}}$.
See Fig.~\ref{fig:SimpScal}(b).

With more effort an analoguous calculation can be carried through for the case of $R$ even and $J_K'=0$. One finds~\cite{Sorensen07b}:
\begin{equation}
S_{\mathrm{imp}}(J^{\prime}_K=0,r,\RR)=(1-p)\ln(2)=S_{\mathrm{imp}}^{\mathrm{IVB}}, \RR\ \mathrm{even}.  \label{eq:sfpeven}
\end{equation}
We see that in this case there is  no contribution from
the single particle entanglement as one would expect since there are no solitons (single particles) present in the ground-state.
Instead, the impurity entanglement is given purely by the impurity valence bond 
impurity valence bond (IVB) (See section~\ref{sec:boundent}) where in the present case one 
identifies the probability that the IVB does not cross the boundary
between regions $A$ and $B$ with the probability of finding a soliton in region $A$. 
See Fig.~\ref{fig:SimpScal}(a).

Some surprising observations can be found by performing numerical calculations of $S_{\mathrm{imp}}$ away from
the MG point where the above variational calculations are quasi-exact. One finds~\cite{Sorensen07b}:
(i) $S_{\mathrm{imp}}$ for both even and odd $R$ is non-zero over the entire range of $r$ and is not limited
by the correlation length which is effectively zero.
(ii) $S_{\mathrm{imp}}(J_K'=0,r,R)$ for $R$ even changes only slightly when $J_2$ is decreased from $J_2=J/2$ to
$J_2=J^2_c$ and for $J_2\le J_2^c$ in the gapless phase it appears not to change at all with $J_2$.
In all cases, $0\le J_2 \le J/2$, does the IVB picture seem to correctly describe $S_{\mathrm{imp}}$. This is
illustrated in Fig.~\ref{fig:SimpScal}(a). IT is perhaps surprising that the IVB picture works relatively well
for $J_2<J_2^c$ where long-range valence bonds are present in the ground-state.
(iii) $S_{\mathrm{imp}}(J_K'=1,r,R)\sim S_{\mathrm{imp}}^{\mathrm{SPE}}$ for $R$ odd decreases rapidly with $J_2$ from $J/2$ to $J_2^c$
where it vanishes. For $J_2\leq J_2^c$ this part of the impurity entanglement is negligible. Hence,
it seems likely that this fact is related to the system becoming gapless at $J_2^c$. This is
illustrated in Fig.~\ref{fig:SimpScal}(b).

Finally we note that, the concurrence of the end spins in the dimerized $J_1-J_2$ $s=1/2$ model have been studied~\cite{Venuti06}.

\section{Conclusions}
We have reviewed recent results related to the impurity contribution to the entanglement, $S_{\mathrm{imp}}$, 
arising from a quantum impurity or boundary. Most notably it has by now clearly been established that the entanglement
can be dramatically changed by the presence of an impurity even in the case where the physical coupling to the impurity
is zero. The role played by different boundary conditions in $1+1$ dimensional critical systems is well understood
and in agreement with results from CFT. However, for the case of mixed boundary conditions (section~\ref{sec:boundent}) 
quantitative theory is not yet available and a detailed understanding of the $R$ dependence of bulk impurity
effects (section~\ref{sec:peschel}) would be desirable. A consistent picture of the
entanglement of a qubit with the environment as described
by the spin-boson problem and the entanglement arising from Kondo impurities has been developed based on
established theory of Kondo systems. A more heuristic picture of impurity entanglement in gapped (and to a certain
extent also critical one dimensional systems) based on valence bond physics and matrix product states is emerging.
Useful notions of single particle entanglement (SPE) and impurity valence bonds (IVB) have been introduced.
Some details of this picture are still missing. As an example,
how the single particle entanglement disappears as the system approaches criticality is still an open problem.
Comparatively few results are available for quantum impourity entanglement in two (and higher) dimensional
quantum systems and a detailed theory is still lacking.
\ack
We are grateful to F. Alet, S. Capponi, M.-S. Chang, J. Cardy and  M. Mambrini for interesting discussions. This research was
supported by NSERC (all authors), the CIFAR (IA) CFI (ESS). 
NL thanks LPT (Toulouse) for hospitality. Part of the numerical simulations have been performed on the WestGrid network.
This work was made possible by the facilities of the Shared Hierarchical Academic Research Computing Network (SHARCNET:www.sharcnet.ca).

\vspace{15mm}
\bibliographystyle{unsrt}
\bibliography{qie_rev}
\end{document}